\documentclass[acmsmall]{acmart}
\AtBeginDocument{%
  \providecommand\BibTeX{{%
    \normalfont B\kern-0.5em{\scshape i\kern-0.25em b}\kern-0.8em\TeX}}}

\usepackage{epstopdf}
\usepackage{ifpdf}
\usepackage{diagbox} 
\usepackage{booktabs} 
\usepackage{enumerate}
\usepackage{subfigure}
\usepackage{verbatim}
\usepackage{balance}
\usepackage{comment}
\usepackage{amsmath}
\usepackage{autobreak}
\usepackage{enumitem}
\usepackage{booktabs}
\newcommand{\tabincell}[2]{\begin{tabular}{@{}#1@{}}#2\end{tabular}}
\usepackage{algorithm}
\usepackage{graphicx}
\usepackage{algorithm,algorithmic}
\usepackage{lipsum}
\usepackage[mathlines,switch]{lineno}

\def\BibTeX{{\rm B\kern-.05em{\sc i\kern-.025em b}\kern-.08em
    T\kern-.1667em\lower.7ex\hbox{E}\kern-.125emX}}
\usepackage{multirow}
\usepackage{multicol}
\usepackage{tabularx} 
\usepackage{tabu}
\usepackage{threeparttable}
\usepackage{color}
\usepackage{url}
\usepackage{anyfontsize}
\usepackage[utf8]{inputenc} 
\usepackage{setspace}
\usepackage{soul}  
    \soulregister{\cite}7
    \soulregister{\citep}7
    \soulregister{\citet}7 
    \soulregister{\ref}7
    \soulregister{\pageref}7
\usepackage{tcolorbox}

\newtcolorbox{summarybox}[1][]
{
    sharp corners,
    left=1mm,
    right=1mm,
    boxrule=0.3mm,
    colback=yellow!30!white
    #1,
}

\begin{document}

\title{Systematic Literature Review of Commercial Participation in Open Source Software}

\author{Xuetao Li}
\email{xuetaoli@bit.edu.cn}
\affiliation{%
  \institution{Beijing Institute of Technology}
  \city{Beijing}
  \country{China}
}

\author{Yuxia Zhang}
\email{yuxiazh@bit.edu.cn}
\affiliation{
  \institution{Beijing Institute of Technology}
  \city{Beijing}
  \country{China}
}

\author{Cailean Osborne}
\email{cailean.osborne@oii.ox.ac.uk}
\affiliation{
  \institution{Oxford Internet Institute, University of Oxford}
  \city{Oxford}
  \country{United Kingdom}
}

\author{Minghui Zhou}
\email{zhmh@pku.edu.cn}
\affiliation{%
  \institution{School of Computer Science, Peking University}
  \city{Beijing}
  \country{China}
}

\author{Zhi Jin}
\email{zhijin@pku.edu.cn}
\affiliation{%
 \institution{School of Computer Science, Peking University}
 \city{Beijing}
 \country{China}
}

\author{Hui Liu}
\email{liuhui08@bit.edu.cn}
\affiliation{
  \institution{Beijing Institute of Technology}
  \city{Beijing}
  \country{China}
}

\renewcommand{\shortauthors}{Xuetao and Yuxia, et al.}

\begin{abstract}
Open source software (OSS) has been playing a fundamental role in not only information technology but also our social lives. Attracted by various advantages of OSS, increasing commercial companies take extensive participation in open source development and have had a broad impact. This paper provides a comprehensive systematic literature review (SLR) of existing research on company participation in OSS. We collected 92 papers and organized them based on their research topics, which cover three main directions, i.e., participation motivation, contribution model, and impact on OSS development. We found the explored motivations of companies are mainly from economic, technological, and social aspects. Existing studies categorize companies' contribution models in OSS projects mainly through their objectives and how they shape OSS communities. Researchers also explored how commercial participation affects OSS development. We conclude with research challenges and promising research directions on commercial participation in OSS. This study contributes to a comprehensive understanding of commercial participation in OSS development. 
\end{abstract}




\maketitle

\section{Introduction} \label{sec:introduction}

In the past two decades, OSS has been widely used in software production, and a large scale of companies are rooted in OSS projects \cite{goggins2021making}. It is reported that more than 90\% of software products contain OSS components \cite{harutyunyan2020managing, o2021coproduction}. Enterprises have recognized the potential of OSS, and their involvement has had a broad and deep impact on OSS communities \cite{A.D.-166, RoblesDuenas-177}. As reported, companies contribute more than 90\% of the code in each release of OpenStack, an OSS cloud computing system \cite{zhang2019companies}. Researchers refer to OSS with extensive commercial participation\footnote{In this paper, commercial participation in OSS is used to demonstrate the phenomenon where companies have assigned employees or hired volunteers to make contributions to an OSS project. This paper will use company, enterprise, and commercial organization indiscriminately.} as OSS 2.0 \cite{fitzgerald2006transformation}.

The cathedral and the bazaar \cite{2001The} are used as a metaphor to discuss two fundamentally different development styles, i.e., the ``cathedral'' model of most companies versus the ``bazaar'' model of open source. In the past 20 years, rich merchants from the cathedral, i.e. companies, have joined the development of OSS projects and have been playing an increasingly important role. 
This influx of commercial participation, in stark contrast to the organic and grassroots nature of the bazaar \cite{2001The}, holds the power to shape the very trajectory of OSS projects \cite{zhang2018companies}. Companies may either dominate the development of OSS projects to maximize their benefits \cite{ZhangStol-196}, or directly withdraw from the OSS projects once their business strategies change \cite{ZhangLiu-191}. The essential differences between companies and open source, e.g., profit-driven motives \textit{vs.} open collaborative spirit, bring both challenges and risks to the sustainability of OSS projects. 

This phenomenon has attracted considerable research and discussion. For instance, Zhang et al. provided a landscape of commercial participation in OpenStack \cite{Y.H.-30, zhang2018companies, ZhangStol-196}. Van der Linden et al. reviewed how large European companies can navigate OSS projects \cite{van2009commodification}. Riehle presented a series of studies discussing open source business model \cite{Riehle-48, D.-150, Riehle-153, Riehle-157, Riehle-186}. Gonzalez-Barahona et al. discussed how companies interact with free software communities \cite{gonzalez2013understanding}. However, current studies on OSS are scattered and have different focuses, and companies and OSS communities still use their empirical experience as a guide to interacting with each other rather than a systematic architecture \cite{west2008designing}.

A few studies have conducted literature surveys about commercial participation in OSS, which either have a specific emphasis or have been published more than a decade ago. For example, Wang et al. reviewed the motivations of both individuals and companies for participating in open source projects \cite{Fei-RongDan-128}. Höst et al. \cite{Host-77} conducted a systematic review of how OSS can be deployed in commercial software. A comprehensive understanding of commercial participation in OSS projects is still left as a knowledge gap. Thus, this survey aims to summarize the existing literature to understand commercial participation in OSS comprehensively. The results of this literature review can provide OSS communities, companies, and researchers with a comprehensive perspective of OSS development with the companies involved.



     
We started the systematic literature review of commercial participation in OSS by choosing three digital libraries to gather related articles, i.e., ACM Digital Library, IEEE Digital Library, and Scopus, and employed Google Scholar to broaden our search using a snowball sampling method. Our search terms revolved around OSS, Company, Participant, and their respective equivalents. After conducting inclusion and exclusion filtering, we collected 92 papers published in 51 venues. We analyzed the publication trend, publication venue distribution, and method types of the 92 papers. We found that researchers have explored commercial participation in OSS since 2002 with a small number of papers, and then academic interest in this topic had a steady increment. Twenty-nine percent of the included papers were published in the premier software engineering conferences or journals. Additionally, most papers conducted case studies by quantitatively or qualitatively analyzing the repository data of OSS projects.

This survey mainly stands on the perspective of OSS development, to analyze and discuss the 92 papers from three well-designed dimensions, i.e., companies' motivations for joining OSS development, their participation models in OSS projects, and their impact on OSS development. The reasons for the three dimensions are as follows: companies' \textit{motivations} can profoundly shape their subsequent contributions \cite{zhou2016inflow}; \textit{participation models} can reflect how they interact with OSS communities; the \textit{impact} companies may bring to OSS development is of great importance because it can largely determine the sustainability of both OSS projects and commercial participation. Through content analysis of the 92 papers, we embarked on an exploration of the multifaceted relationship between companies and OSS projects. 
Drawing inspiration from existing literature \cite{Fei-RongDan-128}, we categorized companies' motivations into three distinct types: Economic motives, Technological motives, and Social motives. Each motive type also has subcategories. 
We conducted a thorough analysis of companies' participation models, encompassing their business objectives and contribution performance. Additionally, we also categorized the various collaboration patterns exhibited by multiple companies within the same OSS community.
Furthermore, we endeavored to unravel the \textit{impact} that companies have had on OSS development, including the effects on companies,  volunteers, and OSS projects.

The significance of our survey lies in the following contributions:
\begin{itemize}
    \item We conducted a detailed examination of 92 relevant studies, shedding light on publication trends, the distribution of publication venues, and types of used methods. 
    \item With a comprehensive analysis, we sought to gain an in-depth understanding of companies' diverse motivations, participation models, and impact on OSS development.
    \item By identifying distinct challenges and untapped research opportunities, we aim to facilitate the sustainable development of OSS projects with the participation of companies.
\end{itemize}

This review offers a valuable resource for practitioners from OSS communities or companies, providing an evidence-based understanding of the benefits and engagement policies associated with participation in OSS. For researchers, this review serves as a roadmap, highlighting existing empirical assessments of commercial involvement in OSS communities, while also identifying areas that have yet to be explored or require further investigation.

The remaining sections of the paper are organized as follows: Background on open source software and commercial participation is presented in Section \ref{background}. In Section \ref{sec: methodology}, we introduce the research questions and describe the paper collection methodology, including the search strategy and inclusion/exclusion criteria. Sections \ref{rq1} - \ref{sec:RQ4} introduce the results. We analyze research opportunities in Sections \ref{sec:opportunities} and limitations in \ref{sec:limitations}. Section \ref{sec: relatedwork} discusses the related work, and the conclusions are presented in Section \ref{sec: conclusions}.

\section{Background} \label{background}

Open source software is available with its source code under a license that allows users to study, change, and distribute the software to anyone and for any purpose \cite{2001The, dahlander2008firms, S.J.-104}. OSS is famous for its transparency, collaboration, cost savings, flexibility, and innovation \cite{laurent2004understanding, marsan2012adoption}.

Over time, two main ideologies have emerged in OSS communities: one proposed by the Free Software movement and represented by the Free Software Foundation (FSF), and the other proposed by the Open Source movement and represented by the Open Source Initiative (OSI) \cite{DanielMaruping-56}. The concepts of free software and open source are two closely related but distinct approaches to software development. 

\begin{itemize}
    \item The FSF is a non-profit organization dedicated to promoting and defending the principles of software freedom through the use of free software licenses and advocacy \cite{stallman2003free}. Free software, as championed by the FSF, emphasizes the freedom of users to use, study, modify, and distribute software. The FSF promotes the use of licenses such as the GNU General Public License (GPL), which ensures that users have these freedoms. The focus is on the ethical and social implications of software, advocating for user rights and community collaboration \cite{williams2010free}.
    \item The OSI is an organization that promotes and advocates for the benefits of open source software, encouraging transparency, collaboration, and the use of open source licenses \cite{o2007governance}. Open source software, represented by the Open Source Initiative, emphasizes the practical benefits of making source code openly available. Open source software encourages collaboration, transparency, and community development. The Open Source Initiative promotes licenses such as the MIT License and the Apache License, which allow for greater flexibility and commercial use \cite{DanielMaruping-56}.
\end{itemize}

While there are differences in philosophy and focus between the free software and open source movements, they share common goals of promoting access to software and encouraging collaboration. Both approaches have had a significant impact on the software industry and have led to the development of many widely used software projects and communities.


Commercial entities have increasingly become involved in OSS development as it has grown in popularity and become 
mainstream in information technology. These entities may contribute to OSS projects in a variety of ways, such as funding development, contributing code, and providing resources. 
There are two main forms of the relationship between 
companies and 
OSS communities:

\begin{itemize}
    \item \textbf{Community Open Source} refers to the OSS projects where their community has the autonomy to develop the software. In this relationship, companies are more like organizational contributors to the OSS community, 
    rather than the software being owned by a single company. In the case of the Apache web server\footnote{https://apache.org/}, individual developers, the committers in particular, make decisions about the software development, not a specific company.
    \item \textbf{Commercial Open Source} refers to OSS projects that are owned by a company \cite{Riehle-157}. The direct benefits to the community are captured by the company, which owns the copyright and determines what is accepted into the software code base and what is implemented next. For example, the company Oracle is in charge of the development of MySQL database\footnote{https://www.mysql.com/}.
\end{itemize}

The difference between Community Open Source and Commercial Open Source is not whether it is profitable. As Riehle says \cite{D.-150}: \textit{``community open source can also be commercial}. 
The key distinction is whether the power to make decisions about the project is in the hands of the OSS community or a single entity such as a company. 

Increasingly companies have been involved in OSS projects since 2001 \cite{van2009commodification, lerner2001open}. With the deepening of business involvement, many articles refer to it as the new stage of OSS. For instance, Fitzgerald et al. refer to OSS with commercial participation as OSS 2.0 \cite{fitzgerald2006transformation}, and Watson et al. refer to it as the second generation of OSS \cite{watson2008business} or professional OSS in contrast to the three types of first generation OSS: community OSS, sponsored OSS, and corporate deployment. Letellier et al. treat OSS with commercial involvement as the third OSS generation from the perspective of the organizational structure of OSS communities
\cite{letellier2008open}.

OSS has indeed become an essential part of the software development landscape, and its popularity has been growing steadily. The involvement of commercial entities in OSS development has brought both opportunities and challenges. On the one hand, commercial entities can provide valuable resources and funding to OSS projects, which can help accelerate their development and adoption. On the other hand, the involvement of commercial entities can also lead to conflicts of interest, as they may prioritize their own business objectives over the needs of the community. In any case, it is important to strike a balance between the interests of the community and the interests of commercial entities. That is why commercial participation in OSS has attracted the interest of both academics and industry. This study conducts a systematic literature review on existing research about commercial participation in OSS to summarize the current state of understanding of OSS 2.0, identifying gaps and challenges that further research should address, and providing an empirical foundation for companies' decision-making when getting involved in OSS. 


\section{Study Design}\label{sec: methodology}

This study conducts a systematic literature review (SLR) following the guidelines of Kitchenham and Charters \cite{2007Guidelines}. In this section, we specify the research questions we explored, 
provide the protocol used in our literature review, and detail the process of collecting articles and data analysis. 

\subsection{Research Questions} \label{sec: researchquestions}

\begin{figure*}[ht]
\centering 
\resizebox{\textwidth}{!}{
    \centering
    \includegraphics[width=\linewidth]{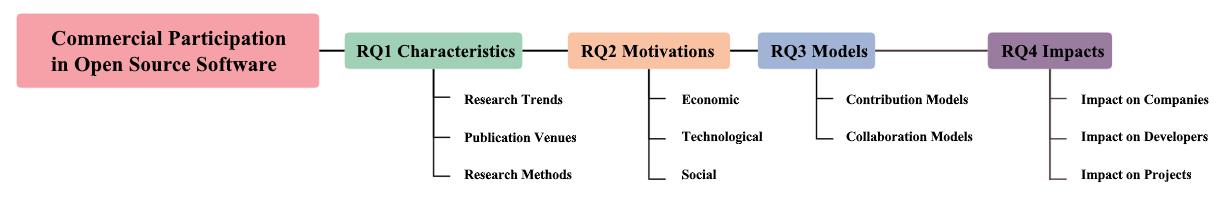}
}
\caption{Mindmap of this SLR}
\label{fig:mindmap}
\end{figure*}

The main objective of this review is to provide a comprehensive understanding and analysis of the entire process of companies' involvement in OSS projects, and to do so, 
we formulate the following three research questions, as shown in Figure \ref{fig:mindmap}: 


We start by understanding some general characteristics of prior studies that have explored commercial participation in OSS, i.e., the research trend, publication venue distribution, and their applied methods.
\begin{itemize}
    \item \textbf{RQ1: What are the characteristics of studies about commercial participation in OSS?} 
    \begin{itemize}
        \item \textbf{RQ1.1: What are the trends of studies on commercial participation in OSS over time?}
        \item \textbf{RQ1.2: What is the distribution of the publication venues?}
        \item \textbf{RQ1.3: What is the distribution of the research methods?}
    \end{itemize}
\end{itemize}

Next, we follow the process of companies' participation in OSS and start from the motivations of their involvement in OSS projects. 
\begin{itemize}
    \item \textbf{RQ2: What are the motivations of for-profit companies for participating in OSS projects?}
    
\end{itemize}

After knowing why companies get involved in OSS communities, we proceed to categorize the models of how companies engage in OSS development.

\begin{itemize}
    \item \textbf{RQ3: What are the participation models of companies in OSS projects?}
    
\end{itemize}

In this paper, we use participation models to present the patterns of how companies interact with the OSS community. With the increasing involvement of companies in OSS, many paradigms, formats, and habits of commercial participation have gradually emerged, which we follow existing work \cite{zhang2019companies} and define as participation models. 

In the last, we focus on an important aspect, i.e., how companies influence the development of OSS projects, which is critical for the sake of OSS sustainability considering the different nature of the bazaar and cathedral and the possible conflict when they run into each other.

\begin{itemize}
    \item \textbf{RQ4 (Impacts): What are the impacts of commercial participation on the development of OSS projects?} 
    
\end{itemize}

\subsection{Study Selection} \label{sec: papercollection}
    
We gathered pertinent literature on commercial involvement in OSS to address our research questions. In this section, we provide a detailed account of the study selection process, including the source of information, the design of study search strings, snowball sampling, and criteria for study selection. 



\subsubsection{Sources of Information}
In this paper, we selected three widely used digital libraries for our initial collection: 
\begin{itemize}
    \item ACM Digital Library\footnote{\url{https://dl.acm.org/}}. This library contains a wide range of ACM electronic publications, including journals and magazines, ACM conference proceedings, and more than 250,000 pages of full-text materials.
    \item IEEE Digital Library\footnote{\url{https://ieeexplore.ieee.org/}}. The IEEE database provides full-text information on journals, proceedings, and current standards collected since 1988.
    \item Scopus\footnote{\url{https://www.scopus.com/}}. This source is the world's largest abstract citation database, covering the world's broadest coverage of peer-reviewed research literature in science, technology, medicine, and the social sciences. 
\end{itemize}   
The three selected information sources support our use of boolean expressions for retrieval. The search process involved two steps. 
First, we searched the three digital libraries. Second, we applied snowball sampling \cite{wohlin2014guidelines} on the citations and authors of the selected papers to complete the search process. Since the reference lists of most of the articles do not give specific links to the articles, we used Google Scholar to snowball sample papers to satisfy the accuracy and efficiency. 
        
\subsubsection{Paper Selection} \label{sec: paperselectionsummary}

\begin{table*}[ht]
    \doublespacing
    \centering
    \caption{Synonym list we used to search related papers.}
    \resizebox{\textwidth}{!}{
    \begin{tabular}{c|c|c}
    \hline
        \textbf{Keyword} & \textbf{Location} & \textbf{Synonyms} \\  \hline
        Open Source Software & Title, Keywords & \tabincell{c}{OSS, Open Source, Free Software, FLOSS, FOSS, Open Source Community, \\Open Source Ecosystems, Open Source Software Ecosystems} \\ \hline
        Company & Title, Keywords	& \tabincell{c}{Commercial, Business, Firm, Corporation, Association, Syndicate, \\Establishment, Organization} \\ \hline
        Participant & Abstract, Keywords & \tabincell{c}{Make use of, Joining, Contributing, Interact,  Inflow, reciprocity\\Retention, Collaborate, Affiliate, Enroll, Participation, corporate} \\ \hline
    \end{tabular}
    }
    \label{tab: Synonym List}
\end{table*}

We used the list of synonyms in Table \ref{tab: Synonym List} to build our search expression. The keyword selection is based on the PICO components: population, intervention, control, and outcome \cite{biolchini2005systematic}. And the synonym selection is based on the guideline of being big and comprehensive. To accurately screen the target papers, we define the search scope of important nouns in the scope of title, abstract, and keywords. In the end, combined with the synonym list we reached the following boolean expression to search the selected database.

\label{alg:algorithm-label}
\begin{algorithmic}
\item[·] Title or Keywords: (``Open Source Software'' or ``OSS'' or ``Open Source'' or ``Free Software'' or ``FLOSS'' or ``FOSS'' or ``Open Source Community'' or ``Open Source Ecosystems'' or ``Open Source Software Ecosystems'') \\

\item[·] Title or Keywords: (``Company'' or ``Commercial'' or ``Business'' or ``Firm'' or ``Corporation'' 
OR ``Association'' or ``Establishment'' or ``Syndicate'' or ``Organization'') 

\item[·] Abstract or Keywords: (``Participant'' or ``Make use of'' or ``Joining'' or ``Contributing'' or ``Interact'' or ``Inflow'' or ``Retention'' or ``Collaborate'' or ``Affiliate'' or ``Enroll'' or ``Participation'') 
\end{algorithmic}

In this step, we selected 129 papers, where 65 papers are from Scopus, 50 papers are from ACM, and 14 papers are from IEEE digital libraries. 

\subsubsection{Snowball Sampling} \label{sec: snowballing}

After the initial analysis, we used one-step snowball sampling \cite{steinmacher2015systematic} to increase the number of papers selected. The snowball sampling includes two steps: 
\begin{enumerate}
    \item Author snowball sampling. We checked the authors' profiles of the initial 129 papers in Google Scholar to see if any relevant papers were not included. 
    \item Backward snowball sampling. We analyzed the cited literature of the selected papers, and relevant literature is included. 
\end{enumerate}
As a result, we got 21 papers by author snowball sampling and 57 papers by backward snowball sampling.

\subsubsection{Selection Criteria} \label{sec: selectioncriteria}
With the 207 retrieved papers, we designed the following inclusion and exclusion criteria to filter out unqualified studies. Specifically, we established three inclusion criteria. 
Papers that meet all of the conditions are initially included:
\begin{itemize}
    \item Written in English;
    \item Must be available as a full paper;
    \item Commercial participation in OSS is (at least a part of) the research, or aspects of commercial participation in OSS are investigated.
\end{itemize}

In terms of studies that were not relevant to the purpose of this review, typical examples of these excluded studies are those that discuss only enterprise behavior that is not related to OSS or only OSS activities regardless of commercial participants. In particular, the following two types of content appear frequently and have been excluded: 

\begin{itemize}
    \item Studies of commercial inner source, which refers to how companies apply open source practices in their inner production environments; we try to limit the scope of this survey from the perspective of OSS communities and leave reviewing inner-source studies as a future avenue.
    \item Studies of the set of processes or tools used by companies to use OSS components as part of their commercial products, while minimizing their risks and maximizing their benefits from such use.
\end{itemize}


We screened all of the selected papers for abstracts and analyzed them for full-text reading. After de-duplication, we performed title, abstract, and keyword analysis, leaving 44 (from Scopus), 19 (from ACM), and 10 (from IEEE) papers. Full paper reading is the last part of the paper selection. In total, 92 papers are included in our system literature review. 
Our paper selection ends in March 2023.








\section{RQ1: What are the characteristics of studies about commercial participation in OSS?} \label{rq1}
In this section, we detail the results arising from the literature review of the three sub-questions of RQ1. 
\subsection{RQ1.1: What are the trends of studies on commercial participation in OSS over time?}

\begin{figure}[ht]
    \centering
    \includegraphics[width=1\linewidth]{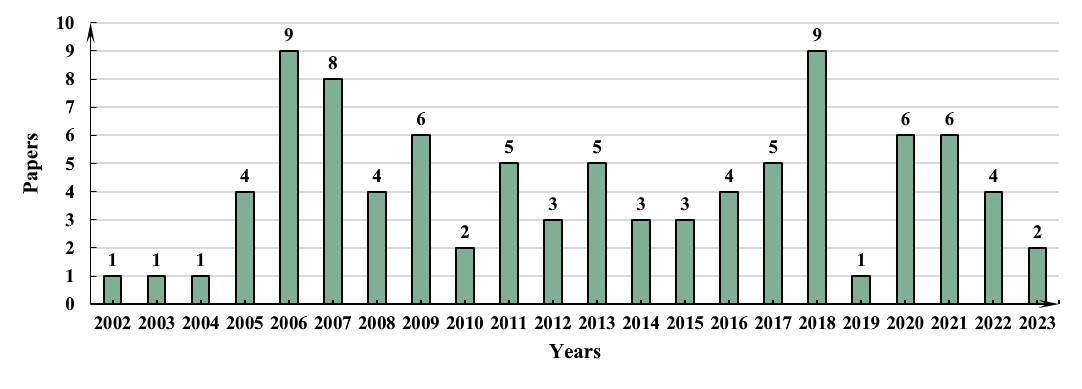}
    \caption{Publication Trend of Studies on Commercial Participation in OSS.}
    \label{tab: year}
\end{figure}

Fig. \ref{tab: year} shows the distribution of 92 studies' publican years. 
We can see that researchers started to explore commercial participation in OSS in 2002, and only a few papers appeared in this early phase. Since 2005, commercial participation in OSS has gained more attention from academia. Compared with other software engineering topics, such as deep learning for software engineering \cite{yang2022survey}, studies on commercial participation in OSS are relatively scattered. With the increasing importance of OSS and the growing number of companies involved, we suggest devoting more research attention to this area.   

\subsection{RQ1.2: What is the distribution of the publication venues?}

We collected each research paper’s publication venue, including conference and journal. The results are shown in Table \ref{Distribution of Publication Venues}. 
Twenty-nine percent of the selected papers were published in the premier conferences or journals of software engineering, such as ICSE, TSE, TOSEM, and FSE.
From Table \ref{Distribution of Publication Venues}, we can see that software engineering is the field in which most papers are published. Besides, there are also 38 studies published in economics, social science, and management venues. Venues outside the computer science category are marked with an asterisk in the table, such as Research Policy and MIS Quarterly. The venue diversity shows the interdisciplinary nature of exploring the phenomenon of commercial participation in OSS.

\begin{table*}
    \centering
    \caption{Distribution of Publication Venues}
    \resizebox{\textwidth}{!}{
    \begin{tabular} {c|c|c}
    \hline
        \textbf{Publication Venues} & \textbf{Count} & \textbf{References} \\ \hline
        International Conference on Software Engineering (ICSE) & 8 &  \cite{A.D.-166} \cite{butler2018investigation} \cite{CapiluppiBaravalle-94} \cite{J.R.-105} \cite{guizani2023rules} \cite{GurbaniGarvert-169}   \cite{Y.X.-195} \cite{Y.M.-190}      \\ \hline
        IFIP International Conference on Open Source Systems (OSS) & 7 &   \cite{BauerHarutyunyan-136}  \cite{daffara2007business} \cite{HaugeSorensen-167} \cite{HaugeZiemer-168} \cite{LindmanJuutilainen-173} \cite{LundellLings-89} \cite{nguyen2011impact}  \cite{RoblesDuenas-177}     \\ \hline
        IEEE Computer & 5 &  \cite{Riehle-48} \cite{D.-150} \cite{Riehle-186} \cite{riehle2020single}  \cite{Y.H.-30}\\ \hline
        Research Policy   & 5 &   \cite{ColomboPiva-72} \cite{dahlander2005relationships}  \cite{OsterlohRota-164} \cite{Stam-80} \cite{West-33}\\ \hline
        IEEE Transactions on Software Engineering (TSE) & 4 & \cite{ButlerGamalielsson-49} \cite{Capra123}   \cite{B.H.-120} \cite{zhang2019companies} \\ \hline
        IEEE Software & 3 & \cite{aberdour2007achieving} \cite{gonzalez2013understanding} \cite{van2009commodification}  \\ \hline
        Information Systems Research (ISR)   & 3 & \cite{FellerFinnegan-149} \cite{HoRai-58} \cite{StewartAmmeter-131} \\ \hline
        MIS Quarterly   & 3 & \cite{DanielMaruping-56} \cite{StewartGosain-155} \cite{agerfalkFitzgerald-180}   \\ \hline
        Academy of Management Proceedings (AOM)   & 2 & \cite{AlexyHenkel-87} \cite{wagstrom2010impact}  \\ \hline
        ACM Transactions on Software Engineering and Methodology (TOSEM) & 2 & \cite{ZhangLiu-191} \cite{ZhouMockus-41}  \\ \hline
        Americas Conference on Information Systems (AMCIS)   & 2 & \cite{LinkQureshiv-61} \cite{Riehle-153}  \\ \hline
        Information and Software Technology (IST) & 2 & \cite{Host-77} \cite{SHAHRIVAR2018202}   \\ \hline
        International Journal of Information Management (IJIM)   & 2 & \cite{Andersen-GottGhinea-27} \cite{linaaker2016firms}   \\ \hline
        International Symposium on Open Collaboration (OpenSym) & 2 & \cite{link2017understanding} \cite{TanshoNoda-97} \\ \hline
        Management Science   & 2 & \cite{EconomidesKatsamakas-43} \cite{von2006promise}  \\  \hline
        The Journal of Strategic Information Systems (JSIS)   & 2 & \cite{marsan2012adoption} \cite{morgan2014beyond} \\ \hline
        Annual Hawaii International Conference on System Sciences (HICSS)   & 1 & \cite{J.S.-31} \\ \hline
        Association for Information Systems (AIS)   & 1 & \cite{SchaarschmidtStol-197} \\ \hline
        Baltic Journal of Management (BJM)   & 1 & \cite{yu2020role} \\ \hline
        Computer Software and Applications Conference (COMPSAC) & 1 & \cite{O.M.-152} \\ \hline
        European Conference on Pattern Languages of Programs (EuroPLoP) & 1 & \cite{Weiss-106} \\  \hline
        First Monday & 1 & \cite{BonaccorsiRossi-46} \\ \hline
        IBM Systems Journal & 1 & \cite{capek2005history} \\ \hline
        iConference & 1 & \cite{JohriNov-95} \\ \hline
        IEEE Intelligent Systems & 1 & \cite{S.G.-127} \\ \hline
        Information and Organization   & 1 & \cite{SchaarschmidtWalsh-66} \\ \hline
        Information Systems and e-Business Management (ISeB)   & 1 & \cite{Riehle-157} \\ \hline
        Information Technology (IT)   & 1 & \cite{gonzalez2013trends} \\ \hline
        International Conference On Information Systems (ICIS)   & 1 & \cite{DanielMaruping-187} \\ \hline
        International Conference on Machine Learning and Cybernetics (ICMLC) & 1 & \cite{Fei-RongDan-128} \\ \hline
        International Conference on Software Business (ICSOB) & 1 & \cite{NguyenDucCruzes-60} \\ \hline
        International Journal of Industrial Organization (IJIO)   & 1 & \cite{LlanesdeElejalde-23} \\ \hline
        International Journal of Information Systems and Management (IJISAM)   & 1 & \cite{GermonprezKendall-182} \\ \hline
        International Journal of Innovation and Technology Management (IJITM) & 1 & \cite{SchaarschmidtVonKortzfleisch-65} \\ \hline
        International Symposium on Wikis and Open Collaboration (WikiSym) & 1 & \cite{CoughlanNoda-96} \\ \hline
        Journal of Enterprise Information Management (JEIM)   & 1 & \cite{mouakhar2017open} \\ \hline
        Journal of Systems and Software (JSS) & 1 & \cite{CapraFrancalanci-78} \\ \hline
        Journal of the Brazilian Computer Society (JBCS)  & 1 & \cite{dias2018drives} \\ \hline
        Knowledge, Technology \& Policy & 1 & \cite{bonaccorsi2006comparing} \\ \hline
        Long Range Planning (LRP)   & 1 & \cite{DahlanderMagnusson-42} \\ \hline
        New Media \& Society   & 1 & \cite{ONeilMuselli-50} \\ \hline
        R\&D Management   & 1 & \cite{WestGallagher-29} \\ \hline
        Software-intensive Business Workshop on Start-ups, Platforms and Ecosystems (SiBW) & 1 & \cite{SpijkermanJansen-184} \\ \hline
        South African Institute of Computer Scientists and Information Technologists (SAICSIT) & 1 & \cite{munga2009adoption} \\ \hline
        The ACM International Conference on the Foundations of Software Engineering (FSE)\footnote{Formerly named ``The ACM Joint European Software Engineering Conference and Symposium on the Foundations of Software Engineering (ESEC/FSE)''} & 1 & \cite{ZhangStol-196} \\ \hline
        The Proceedings of the ACM on Human Computer Interaction (HCI) & 1 & \cite{GuizaniChatterjee-108} \\ \hline
        Web Science (WebSci) & 1 & \cite{HomscheidSchaarschmidt-100} \\ \hline
        Workshop on Gender Equality in Software Engineering (GE)  & 1 & \cite{newtonleveraging} \\ \hline

        Others & 4 &  \cite{BaranesVuong-54} \cite{birkinbine2020incorporating}  \cite{o2021coproduction} \cite{snarby2013collaboration} \\ \hline
        
    \end{tabular}
    }
    \label{Distribution of Publication Venues}
\end{table*}

\subsection{RQ1.3: What is the distribution of the research methods?}

To answer this question, the first two authors of our study analyzed every inclusion paper's research methods based on the full paper reading.
The classification standard of the methods is to give priority to the research methods clearly stated by the authors. For studies that have not clearly stated their applied methods, we extracted and categorized their methods by reading their method sections. 

\begin{figure}[ht]
\includegraphics[width=.5\linewidth]{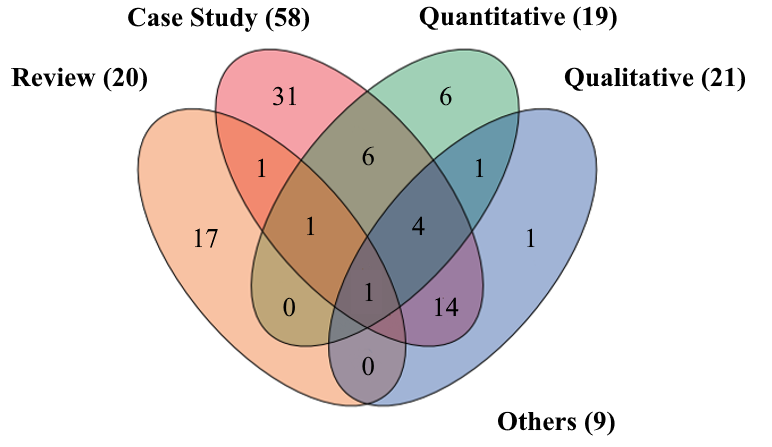}
\caption{Distribution of Reported Methods in the Selected Studies.}
\label{tab: distribution}
\end{figure}

Fig. \ref{tab: distribution} shows the method distribution of the 92 selected studies. More than one method may be used when conducting the same study, which is presented by the intersections in Fig. \ref{tab: distribution}. 
We can see that most of the papers reported case studies and gave a quantitative or qualitative analysis of collected data. Specifically, 58 out of the 92 studies (63.0\%) used the case study method to explore how companies participate in OSS. Twenty studies (21.7\%) are systematic literature reviews. Nineteen studies (20.7\%) use the quantitative method, and 21 papers (22.8\%) use the qualitative method.

\section{RQ2: What are the motivations of for-profit companies for participating in OSS projects?} \label{sec:RQ2}
There is a paradox, i.e., \textit{why do companies want to share their resources even though they are less likely to dominate the market and their competitors can acquire the same resources?} The content of this section can answer this question, i.e., what motivates companies to participate in OSS development? After analyzing the selected studies, we find that social, technological, and economic are the main aspects that companies focus on. Economic motivations are the principal reason for commercial participation \cite{Fei-RongDan-128}, and social and technological incentives can also lead to economic promotion. Table \ref{tab: motivations} provides a comprehensive list of these identified motivations from the three categories and their references. We analyze each category's motivations in the following subsections.




\begin{table*}[ht]
    \centering
    \caption{List of Companies' Motivations toward Participating in OSS Development}
    \resizebox{\textwidth}{!}{
\begin{tabular}{ll|l}
\hline
\multicolumn{2}{c|}{\textbf{Motivation Field}} &
  \multicolumn{1}{c}{\textbf{Reference}} \\ \hline
\multicolumn{1}{l|}{\multirow{5}{*}{Economic}} &
  More Effective Marketing &
  \cite{CapraFrancalanci-78} \cite{Riehle-48} \cite{gruber2006new} \cite{HaugeSorensen-167} \cite{Riehle-153} \cite{Riehle-157}   
    \\ \cline{2-3} 
\multicolumn{1}{l|}{} &
  Distribution Model Change & \cite{BonaccorsiRossi-46}
  \cite{Fei-RongDan-128} \cite{gonzalez2013trends} \cite{morgan2014beyond} \cite{D.-150} \cite{Riehle-157} \cite{west2006challenges}      \\ \cline{2-3} 
\multicolumn{1}{l|}{} &
  More Customers & \cite{Andersen-GottGhinea-27} \cite{Fei-RongDan-128} \cite{HaugeSorensen-167} \cite{D.-150} \cite{Riehle-157}    \\ \cline{2-3} 
\multicolumn{1}{l|}{} &
  Cost Reduction &
  \tabincell{c}{ \cite{Andersen-GottGhinea-27} \cite{bonaccorsi2006comparing} \cite{dahlander2005relationships} \cite{Fei-RongDan-128} \cite{O.M.-152}  \cite{HaugeSorensen-167} \cite{hippel2003open}  \cite{LindmanJuutilainen-173} \cite{LlanesdeElejalde-23} \cite{morgan2014beyond} \cite{D.-150} \cite{Riehle-153} \cite{Riehle-157}          \cite{agerfalkFitzgerald-180}  } \\ \cline{2-3} 
\multicolumn{1}{l|}{} &
  Market Access & \cite{ColomboPiva-72}
  \cite{Riehle-48} \cite{dahlander2005relationships} \cite{Fei-RongDan-128} \cite{O.M.-152} \cite{LlanesdeElejalde-23} 
  \cite{morgan2014beyond} \cite{D.-150}        \\ \hline
\multicolumn{1}{l|}{\multirow{3}{*}{Technological}} &
  Higher Quality and Higher Availability &
  \tabincell{c}{\cite{BonaccorsiRossi-46} \cite{bonaccorsi2006comparing} \cite{capek2005history} \cite{Fei-RongDan-128} \cite{O.M.-152}  \cite{HaugeSorensen-167} \cite{LindmanJuutilainen-173} \cite{morgan2014beyond} \cite{Riehle-157} \cite{west2006challenges} } \\ \cline{2-3} 
\multicolumn{1}{l|}{} &
  Better Access To Requirement &
    \cite{Andersen-GottGhinea-27}  \cite{dahlander2005relationships}  \cite{Fei-RongDan-128} \cite{guizani2023rules} \cite{Riehle-153} \cite{Riehle-157} \\ \cline{2-3} 
\multicolumn{1}{l|}{} &
  Driving Innovation &
  \tabincell{c}{\cite{Andersen-GottGhinea-27} \cite{BonaccorsiRossi-46} \cite{bonaccorsi2006comparing} \cite{ColomboPiva-72} \cite{Fei-RongDan-128} \cite{guizani2023rules} \cite{morgan2014beyond} \cite{Riehle-186} \cite{west2006challenges}        \cite{agerfalkFitzgerald-180}      \cite{WestGallagher-29} } \\ \hline
\multicolumn{1}{l|}{\multirow{4}{*}{Social}} &
  Building Credibility \& Trust & \cite{guizani2023rules}  \cite{LindmanJuutilainen-173} \cite{morgan2014beyond} \cite{west2006challenges}   \cite{agerfalkFitzgerald-180}  \\ \cline{2-3} 
\multicolumn{1}{l|}{} &
  A Larger Labor Pool & \cite{bonaccorsi2006comparing}
  \cite{Fei-RongDan-128} \cite{guizani2023rules} \cite{morgan2014beyond} \cite{mouakhar2017open} \cite{D.-150} \cite{Riehle-153}        \cite{agerfalkFitzgerald-180}\\ \cline{2-3} 
\multicolumn{1}{l|}{} &
  Improving Reputation & \cite{bonaccorsi2006comparing} \cite{ColomboPiva-72}  \cite{Riehle-48} \cite{FellerFinnegan-149}  \cite{guizani2023rules}  \cite{mouakhar2017open}   \cite{agerfalkFitzgerald-180}  \\ \cline{2-3} 
\multicolumn{1}{l|}{} &
  Promoting Standardization & \cite{dahlander2005relationships}
  \cite{Fei-RongDan-128}  \cite{LindmanJuutilainen-173}    \cite{WestGallagher-29} \cite{west2006challenges} \\ \hline
\end{tabular}}
\label{tab: motivations}
\end{table*}

\subsection{Economic Motivations} \label{sec:RQ2-economic}
After the full-reading analysis, we found twenty studies have proposed five economic types of companies' motivations towards participating in OSS development.  

\begin{itemize}
\item[\textsc{E1}] \textbf{More Effective Marketing.} Economic efficiency is the fundamental starting point of all activities of enterprises \cite{shatalova2015methodological}. OSS community marketing is usually more effective than the company's own marketing because 
releasing a product in an open-source way can make it available to a large number of users. If the community is happy with the product, it is likely to share its experience with others, thus providing free marketing and increased publicity for the OSS provider \cite{HaugeSorensen-167}. Potential users sometimes prefer to trust OSS community marketing because everyone in OSS communities plays a witness and can learn about the maintenance situation of OSS projects from the code change history \cite{Riehle-153}. This kind of marketing is a free and effective way to win recognition from customers. 
OSS products are highly competitive in the market.
The software product itself is also an advertisement. The free user base is significant because many industrial OSS vendors sell services related to the free product, and more free users mean more potential paid users. Customers can learn more about the company from the software product if it is a commercial OSS project \cite{HaugeSorensen-167}. In addition, Gruber and Henkel gave evidence that 
participating in OSS projects, engaging in discussions, and supporting activities can serve as effective marketing for software firms, as potential buyers often seek capable software firms in relevant OSS projects \cite{gruber2006new}.


\item[\textsc{E2}] \textbf{Distribution Model Changes.} A distribution model outlines how goods make their way from the manufacturer to a consumer outlet. In the traditional closed-source distribution model, companies mainly focus on selling software products \cite{Fei-RongDan-128}. However, when companies provide their products based on open source, they obtain indirect revenue by selling related products 
\cite{west2006challenges, Fei-RongDan-128, Riehle-153}, instead of directly selling products. Researchers found that switching to open source can lead to increased profit per sale and increased sales volume \cite{gonzalez2013trends}. Besides, companies can get to market faster with a superior product at a lower cost than traditional competitors \cite{D.-150, Riehle-153}. 


\item[\textsc{E3}] \label{tab: E3} \textbf{Customer Acquisition.} Customers prefer OSS because the prices are lower than the closed-source software. For commercial companies, more users mean better economics, and switching from closed-source products to OSS results in more potential customers. Especially in commercial open source, a self-supporting community can grow into a large user base that can later be converted into paying customers \cite{D.-150}. In addition, the risk to customers may be lower because code auditing for everyone and addressing security issues are much more efficient than in close source \cite{Riehle-157, Fei-RongDan-128}.

\item[\textsc{E4}] \textbf{Cost Reduction.} With almost zero cost of participation, it is tempting for companies to take advantage of the OSS community \cite{hippel2003open, agerfalkFitzgerald-180}. One way to utilize this is through alternative outsourcing, where an OSS community can provide companies with a continuously maintained and developed product, reducing the corresponding costs \cite{Fei-RongDan-128}. Besides, from the perspective of software development \cite{dias2018drives}, external developers play an important role in OSS development and can decrease a considerable cost for companies. Because the external developers' contributions range from documentation to complex tasks such as reviewing code or developing core code.

Individual companies typically invest less in OSS compared to closed source. However, when multiple companies invest in the same OSS community, the community as a whole receives a greater amount of resources than what a single company would invest in closed source software \cite{LlanesdeElejalde-23}. As companies collaborate more with the community, the cost may vary based on each company's share of contributions to the OSS project. However, this will not reduce the company's profits because it can charge higher prices due to better service and faster version changes \cite{D.-150, Riehle-157, gonzalez2013trends}. In addition, the community can perform testing, maintenance, and bug fixes, thus reducing development costs and improving the software \cite{HaugeSorensen-167}.
    
\item[\textsc{E5}] \textbf{Market Access.} Because of its cost-effectiveness, friendly licensing policy, high quality, and broad user base, OSS has a higher probability of penetrating new markets in their early stages \cite{Fei-RongDan-128}. Typically, commercial open source communities dominate the initial open source market because they can provide a clear direction and mobilize more resources than community open source projects. However, as the OSS project matures, this advantage may diminish because competing community OSS projects can obtain the same tech source and may have access to more volunteer resources. 
Both have better market access, but only with strong intellectual property protection or other competitive advantages can they triumph and sustain a new market \cite{D.-150}. Meanwhile, OSS communities also provide developing companies with an opportunity. Companies can get free resources and achieve business innovation \cite{ColomboPiva-72}. This is valuable because only open source can solve vendor lock-in and achieve innovation \cite{D.-150}.
\end{itemize}

\subsection{Technological Motivations} \label{sec: MotivationsTechnological}\label{sec:RQ2-technological}
We categorized three technological motivations of commercial participation in OSS from sixteen studies, as shown in Table \ref{tab: motivations}.

\begin{itemize}
\item[\textsc{T1}] \textbf{Higher Quality and Higher Availability} are two main attractive characteristics of OSS \cite{M.-28} when compared with traditional software. The two characteristics have a strong relation with rapid requirement acquisition and innovation, which explains why companies would like to share code and knowledge with the OSS community \cite{Fei-RongDan-128}. 
Although the cost of each company has decreased, the total investment received for the whole OSS community has increased because of the breadth of corporate and volunteer participation \cite{O.M.-152}. This also promotes better software quality. According to IBM's open source experience, the open source development style attracted highly skilled developers, and the overlap between developers and users of a given OSS project allowed for excellent and open communication, rapid development cycles, and intensive real-world testing, resulting in software that was often very good and sometimes excellent by IBM's standards \cite{capek2005history}.

\item[\textsc{T2}] \textbf{Better Access To Requirement.} Requirements gathering and analysis are significant but with huge costs. OSS provides an open platform for both customers and developers on which real users 
can observe and engage with the community and discuss product features \cite{Riehle-153}. As a result, the company's products can easily achieve the customers' needs and then benefit from the customers' needs \cite{Riehle-157}.

\item[\textsc{T3}] \textbf{Driving Innovation.} Taking full advantage of OSS projects can help companies innovate, especially small- and medium-sized companies \cite{Fei-RongDan-128, BonaccorsiRossi-46}. The combination of innovation and open source is common and efficient in the real practice of many smaller companies \cite{WestGallagher-29}. OSS offers companies the characteristic of flexibility, which enables free customization, experimentation, and modification for specific needs \cite{morgan2014beyond, agerfalkFitzgerald-180} based on OSS. This kind of innovation can even lead to a change in product direction. In the early days, OSS often appeared as a copycat of commercial software \cite{agerfalkFitzgerald-180}. When the spark of innovation is ignited in OSS communities, several OSS projects have become standards in their fields, such as the Linux kernel \cite{sen2012open}. Existing work discovered that large IT companies with a wide range of products participated in open source because they could not ignore any significant external sources of innovation available to their competitors \cite{WestGallagher-29}. Despite the different starting points of large and small companies - large companies to gain access to the same resources as their competitors, and small companies to realize their own technological innovations - all companies can innovate in the course of OSS cooperation.
\end{itemize}

\subsection{Social Motivations} \label{sec:RQ2-social}
The development of OSS is full of large social collaboration activities because contributors are distributed around the world, and their cooperation is transparent and traceable. 
Some companies do not derive direct economic benefits from OSS community participation but are still willing to allocate technology and human resources to OSS projects. Because OSS involvement facilitates their collaborative networking and creates access to social capital, thereby contributing to financial performance \cite{yu2020role}. Fourteen studies have mentioned four types of social motivations for commercial participation in OSS. 

\begin{itemize}
\item[\textsc{S1}] \textbf{Building Credibility \& Trust.} Due to the transparency of open source, companies' contributions are transparent, allowing the outside world to see the quality and maintenance status of their contributions, which helps establish trust between users and companies \cite{guizani2023rules}. This kind of transparency is also what companies that use open source as an alternative to outsourcing would like to see \cite{agerfalkFitzgerald-180}. 

\item[\textsc{S2}] \textbf{A Larger Labor Pool.} The entire OSS community is a large labor pool for companies. Companies prefer to look for top developers in OSS projects because they can observe and evaluate developers' capability through their commit history \cite{agerfalkFitzgerald-180}. Companies can avoid the risks of inappropriate hiring because people in OSS communities have proven themselves \cite{Riehle-153}. Benjamin Birkinbine argues that the greatest value for companies comes from the processes of OSS development, enabled by a globally distributed and free labor force, rather than the products themselves \cite{birkinbine2020incorporating}. However, a larger pool means easier hiring and firing, so the salary of employees may be lower. To some extent, the reduced costs are passed on to the employees \cite{D.-150, Fei-RongDan-128}.

\item[\textsc{S3}] \textbf{Improving Reputation.} Companies'  reputation can influence the choice of customers and the onboarding of new employees to some extent. Playing an important role in the OSS community is a valid way to improve reputation outside the network \cite{FellerFinnegan-149}.
As for small and medium companies, the acceptance of code into the official version is a qualification of the company's code quality and technology, which improves the company's reputation \cite{ColomboPiva-72}.

\item[\textsc{S4}] \textbf{Promoting Standardization.} This is another answer to why companies contribute to OSS projects. OSS products can attract more users as we discussed in Sec. \ref{tab: E3}, which can help establish their technology as an industry standard. Huge economies of scale will come from the industry standard. This is essential because companies can build a perfect product system around standardization, with great revenue potential and great promotion to the company's reputation \cite{WestGallagher-29, Fei-RongDan-128}. 
\end{itemize}


\begin{summarybox}
{
\textbf{Summary:}
Overall, existing studies have repeatedly pointed out that the use of OSS can bring several benefits to companies, including economic, technological, and social reasons, which can coexist. Companies' ultimate goals are to reduce costs and make better products. Motivation also varies with company size, as companies at different stages of development may be inclined to different dimensions of motivation. 
}
\end{summarybox}

\section{RQ3: What are the participation models of companies in OSS projects? }\label{sec:RQ3} 
Companies contribute to open source for a variety of reasons. As highlighted in \cite{ButlerGamalielsson-49}, the value companies obtain from OSS projects largely dictates their level and form of engagement. While target goals may be similar, contributing models often differ between companies. After a comprehensive analysis of the selected papers, we found that 13 studies explored companies' contribution models mainly from two angles: the contribution models of individual companies and multiple companies' collaboration models.  
In this section, we provide a comprehensive summary of all these categorized contribution and collaboration models of companies. 
\subsection{Contribution Models}\label{6.1} 
Existing studies proposed contribution models to understand how companies behave in the development of OSS projects, which mainly consist of two parts, i.e., the company’s business objectives and its engagement activities. After analysis, we found two main angles that are used to categorize companies' contribution models: how companies benefit from OSS projects and how companies shape OSS communities.   

\textbf{\textit{How companies benefit from OSS projects.}} Companies' objectives towards OSS projects can drive their involvement and guide their actions for achieving the objectives. Most of the studies distinguished companies' involvement in OSS via how the companies make profits based on the OSS projects, and companies with the same objectives show commonalities in their contribution characteristics. 
\begin{itemize}
    \item \textbf{Distributor}. Companies following this model sell subscriptions to software and associated services based on OSS products or components. This model is suitable for complex software with potentially incompatible components. As pointed out by \cite{wagstrom2010impact, linaaker2016firms, zhang2019companies, Y.M.-190}, companies may package and distribute the entire output of the OSS community, such as Linux distributors, or utilize specific components or applications from the OSS community. Businesses focusing on the entire OSS project or specific components can largely determine where a company contributes. 
    
    Studies \cite{wagstrom2010impact, zhang2019companies, Y.M.-190} found that companies aligning with the project community as a whole tend to make substantial and extensive contributions, attract more volunteer developers, and have a favorable reputation. They are less likely to contribute to the heterogeneity of the community compared to component-focused companies. 

    \item \textbf{Offer commercial versions and sell licenses}. While developing the project as open source, companies in this model make profits by offering a commercially licensed version. The commercial version is normally more systematically integrated and easier for users to use. This model benefits from community contributions and encourages users to transition to the commercially licensed version. Both Riehle et al. \cite{Riehle-186} and Daffara \cite{daffara2007business} identified this type. 

    Companies with this objective tend to reach a dilemma \cite{Riehle-186}, i.e. the OSS product must not be complete enough to avoid competition with its commercial version but must be valuable enough to attract users. Therefore, companies mainly contribute to their commercial version. 
     
    \item \textbf{Complementary Services}. Companies in this model generate revenue by offering services to assist users in implementing and maintaining open-source software effectively. They do not sell software licenses but rather provide expertise and support. This model is first proposed in \cite{Riehle-186}, and ``Selection/Consulting Companies'' in \cite{daffara2007business}, ``Service Provider'' in \cite{linaaker2016firms} and ``Specific Services Oriented'' in Zhang et al. \cite{zhang2019companies} are homogeneous.

    In this type, the role companies play is often not provider, but analyst instead \cite{daffara2007business}, and they focus mainly on a few projects closely related to their own business \cite{zhang2019companies} and make non-intense contributions.
    
    \item \textbf{Integration}. Companies following this strategy contribute through plugins or drivers, integrating their offerings with OSS projects. Zhang et al. \cite{zhang2019companies, Y.M.-190} identified this type of commercial involvement in OpenStack, an open source cloud computing system, where companies like Intel pertain to the integration of OpenStack with its existing products or services. ``Product supporter'' proposed by Lin{\aa}ker et al. \cite{linaaker2016firms} also belongs to this category.

    \item \textbf{Usage Oriented}. Like volunteers who make contributions to OSS to `scratch their own itch', companies also participate in OSS communities because of utilizing the OSS products for their production requirements \cite{zhang2019companies, Y.M.-190}. The main goal of this strategy is to reduce costs and increase efficiency by using OSS projects that are freely available and can be customized to meet specific business needs. Lin{\aa}ker et al. proposed ``Platform user'' to refer to companies that use Apache Hadoop to store and process data \cite{linaaker2016firms}. Similarly, Zhang et al. found that users of OpenStack make extensive–but not large by volume–contributions, with the main focus on deployment tools. ``Operational open source'' defined in \cite{morgan2014beyond} also belong to this type.   

    Usage-oriented companies typically engage in a wide range of projects and prioritize the use of specific deployment tools that align with their objectives and requirements \cite{zhang2019companies}. These companies focus on delivering practical solutions and products that cater to the needs of their target users.
\end{itemize}


\textbf{\textit{How companies shape OSS communities.}} To secure lasting revenue streams, companies seek to control the development of OSS projects in which they participate. Existing studies also categorized companies' contribution models through the perspective of how they shape and control the OSS communities. The control level is manifested through various policies and actions. Dahlander et al. categorized how firms interact with the OSS community in three ways \cite{dahlander2008firms}: accessing communities to broaden the resource base; aligning firm strategies with the community; and assimilating communities to integrate and share results. These three ways show three degrees to which the company uses the community, from interaction at the knowledge level to changing the direction of the company's or even the community's development. There is no doubt that the right method will give the company a competitive advantage, so how to acquire, coordinate, and use OSS is important and deserves thorough exploration. Existing studies also revealed the development pattern of commercial software: most innovative software is initially sold as commercial software, with time the software becomes less and less innovative, and the company uses the software as OSS to cut costs and regain power \cite{van2009commodification, BaranesVuong-54}. This development pattern can be treated as a strategy for companies to balance competitiveness, creativity, and cost in community practice.

Dahlander et al. \cite{dahlander2005relationships} and Lundell et al. \cite{lundell2010open} inspired by natural ecosystems and proposed a typology of three different basic models to elaborate how companies interact with the OSS communities they are involved: parasitic, mutualistic or symbiotic, and commensal. The parasitic approach refers to companies that only focus on their own benefits without considering harm to the OSS community. The symbiotic approach entails businesses participating in co-development with the community while respecting OSS norms and values. The commensalism approach enables businesses to gain from communal resources without actively aiding their development. Companies in either symbiotic or commensalism approach tend to respect OSS norms and values and obey licenses, while the former can have a high possibility of influencing the OSS community. These studies also emphasize the continuous nature of these approaches and highlight the need for companies to navigate the delicate balance between openness and control in their interactions with OSS communities.

Yu summarizes the engagement of companies via three dimensions \cite{yu2020role}, i.e., technology giving, technology taking, and social participation. Similarly, Weiss classified according to the strategic importance of open source to the company \cite{Weiss-106}, which includes four levels: use, contribute, champion, and collaborate. The `use' level involves integrating open source components into products. Companies at this level primarily consume and utilize OSS in their products or services. At the `contribute' level, companies not only use open source components but also engage in the development and improvement of these projects. Companies at the champion level initiate their OSS projects. They take on the role of leading and promoting these projects. The collaborative level involves companies collaborating on the creation of shared assets through OSS projects. The goal is to reduce the cost of product development by working together on common projects.

Zhou et al. \cite{zhou2016inflow} categorized three commercial OSS involvement models via whether a company manages the intellectual property (IP) once it gets involved in an OSS project: hosting, supporting, and collaborating. Hosting means that a company has full control over the OSS project. Supporting refers to the situation when a company supports a project, but the project is controlled by another organization. Collaborating means when a company has shared control over the project with other organizations, which is the same as the 'collaborate' level proposed in \cite{Weiss-106}.

\begin{summarybox}
\textbf{Summary:} Existing studies categorized companies' participation in OSS ecosystems via their diverse strategies or aims of controlling OSS projects. Companies utilize various business models, prioritize community involvement in different ways, and contribute to OSS projects to various extents. In summary, their commercial goals influence their actions and policies within the community, shaping the overall dynamics of the participation model.
\end{summarybox}

\subsection{Collaboration Models} \label{sec:RQ3.2}
Besides exploring individual companies’ strategy models, another line of inquiry focuses on understanding how and why companies, including market rivals, jointly develop OSS projects and build collaborations between companies in an OSS community. 

Lin{\aa}ker et al. explored how different kinds of company stakeholders collaborate in the Apache Hadoop ecosystem \cite{linaaker2016firms}. They observed collaborations among all stakeholder categories, except for companies selling Hadoop-based infrastructure and companies that sell Hadoop-related services. Companies selling packaged Apache Hadoop solutions were the most active, engaging in collaborations with other solution providers and various stakeholder categories. Their results suggest that various stakeholders, including those with both competing and non-competing business models, collaborate within the Apache Hadoop ecosystem, revealing the interconnectedness and collaborative nature of those involved. Similarly, Zhang et al. conducted an empirical study of OpenStack and found that there are three recurring collaboration patterns \cite{Y.M.-190}: intentional collaboration, passive collaboration, and isolated participation. Intentional collaboration refers to the phenomenon where companies actively seek out and pursue collaborations. Passive collaboration means that although companies make contributions to the same OSS projects, they are linked together because of their own interests, and there is no obvious coordination between them. Unlike the first two models of collaboration, isolated participation means that companies are the only contributors to an OSS project and these companies do not collaborate with other companies. They concluded that most companies work with others, or even with their competitors. 

Snarby \cite{snarby2013collaboration} explored companies' collaboration from the communication perspective and found three main categories. The first pattern, named ``gatekeeper'', requires a designated representative to navigate the code and information flow within the OSS project. The ``secure channel'' pattern involved using private communication channels to discuss sensitive OSS project-related issues with other company collaborators. Subsequently, the ``open-core collaboration'' pattern was noted, with companies contributing all code they created to the OSS project's public sources. These patterns offer insight into the diverse methods by which commercial firms engage and collaborate with community-based OSS projects. 

Nguyen et al. found six main themes related to co-opetition among firms, which are: Organizational boundary spanning via gatekeepers, Securing communication among actors on firm competitive advantages, Open-core sourcing policy, Business driven filtering of code sharing, Value of social position in OSS community, and Friendly competitiveness \cite{NguyenDucCruzes-60}. 
The first three patterns are almost identical to Snarby's work. The fourth pattern indicates that not all of the companies' code will be contributed back to the community, companies can contribute two kinds of codes, which are OSS core codes and codes that are considered as open and/or standardized, aiming to work together within the community to further develop the code that they have contributed. The fifth pattern shows that companies want to achieve a central position in the community, get more help from other community members, and make it easier to integrate their codes into the project. The last one shows that although competitors in the market may contribute to the same OSS communities, they treat each other as partners or colleagues in the process of contributing. Overall, company interactions in the community involve both competition and collaboration, intending to gain greater competitive strength for the company.

\begin{summarybox} 
\textbf{Summary: } Companies' collaborations are important to the success of OSS projects. Existing studies explored various patterns employed by companies to foster collaborations in different OSS project cases. Some studies emphasized the importance of finding a balance between collaboration and competition. 
It is worth noting that compared to other research areas, studies of company collaboration within OSS projects remain relatively limited.
\end{summarybox}

\section{RQ4: What are the impacts of commercial participation on OSS development?} \label{sec:RQ4}

The pervasive participation of companies in OSS projects can have a huge impact on many OSS stakeholders, including companies,  volunteers, and OSS projects. In this section, we will discuss these impacts explored by existing studies. Table \ref{table:Impact} provides a list of impacts explored by existing studies. 

\begin{table*}[ht]
    \centering
    \caption{List of Commercial Participation's Impact on OSS Development}
    \label{table:Impact}
    \resizebox{\textwidth}{!}{
\begin{tabular}{ll|l}
\hline
\multicolumn{2}{c|}{\textbf{Impact}} & \textbf{References} \\ \hline
\multicolumn{1}{c|}{\multirow{3}{*}{Impact On Companies}} & Promoting the development of companies & \cite{ButlerGamalielsson-49} \cite{ehls2017open}  \cite{linaaker2016firms}  \cite{LindmanJuutilainen-173} \cite{LinkQureshiv-61} \cite{NguyenDucCruzes-60} \cite{Weiss-106} \cite{WestGallagher-29} \cite{west2008designing}   \cite{yu2020role} \cite{yuxia2017openstack}   \\ \cline{2-3} 
\multicolumn{1}{l|}{}                                      & Shaping the open culture in companies                               & \cite{AlexyHenkel-87} \cite{DanielMaruping-187} \cite{GurbaniGarvert-169} \cite{mouakhar2017open}  \cite{D.-150}          \\ \cline{2-3} 
\multicolumn{1}{l|}{}                                      & Attract more company engagement       &  \cite{Andersen-GottGhinea-27} \cite{ButlerGamalielsson-49} \cite{FellerFinnegan-149} \cite{B.H.-120} \cite{TanshoNoda-97}  \cite{yu2020role}            \\ \hline
\multicolumn{1}{l|}{\multirow{2}{*}{Impact On Volunteers}} & Loss of volunteer developers                                & \cite{birkinbine2020incorporating}   \cite{guizani2023rules} \cite{osterloh2002trust} \cite{StewartGosain-155} \cite{zhang2019companies}  \cite{zhou2016inflow}        \\ \cline{2-3} 
\multicolumn{1}{l|}{}                                      & Role conflict for paid developers                               &  \cite{ButlerGamalielsson-49} \cite{DanielMaruping-56} \cite{HomscheidSchaarschmidt-100}  \cite{SchaarschmidtStol-197}        \\ \hline
\multicolumn{1}{l|}{\multirow{2}{*}{Impact On Projects}}   & Increase project visibility                              &  \cite{A.D.-166} \cite{CapraFrancalanci-78}  \cite{LlanesdeElejalde-23} \cite{newtonleveraging}         \\ \cline{2-3} 
\multicolumn{1}{l|}{}                                      & Reduce project openness                             &  \cite{A.D.-166} \cite{CapiluppiBaravalle-94} \cite{Capra123} \cite{newtonleveraging} \cite{ONeilMuselli-50} \cite{D.-150}  \cite{B.H.-120} \cite{SchaarschmidtVonKortzfleisch-65} \cite{SchaarschmidtWalsh-66} \cite{west2008designing}  \cite{Y.H.-30} \cite{ZhangStol-196} \cite{zhou2016inflow}           \\ \hline

\end{tabular}

}
\end{table*}

\subsection{Impact on Companies}

This chapter describes the impact of commercial involvement in OSS projects, either for the company itself or for other companies involved in the same projects. After analyzing the 40 studies, we categorized three aspects of impact on companies: promoting companies' development, shaping companies' open source culture, and attracting more company engagement. Below we introduce how and what impact existing studies have explored in each aspect. 

\begin{enumerate}

\item[\textsc{IC1}] \textbf{Shaping companies' open source culture.} 
Embracing open source can be beneficial for companies seeking to innovate, attract talent, and navigate the evolving digital landscape. However, transitioning from traditional, hierarchical structures to community-driven open source can be challenging. Open source's decentralized and free-spirited nature often clashes with traditional practices that prioritize profit and control. A key hurdle is bridging the cultural gap between companies and OSS communities. Gurbani et al. identified six areas of incompatibility \cite{GurbaniGarvert-169}, i.e., requirements, work assignments, tool compatibility, software architecture, software process, and incentive structure. 

However, bridging this gap can lead to substantial rewards. For example, companies like OpenUpss demonstrate how openness can unlock a competitive edge \cite{mouakhar2017open}. OpenUpss has attracted a dedicated customer base and fueled its technical expertise through its commitment to sharing code and fostering innovation. Their partnerships with research institutions further solidify their position in the OSS ecosystem. In addition, an open source culture can lead to a more engaged and loyal workforce \cite{DanielMaruping-187, D.-150}. When employees are empowered to share ideas and contribute to OSS projects, they tend to experience increased motivation and find their work more meaningful. Alexy et al. examined the acceptance of OSS by staff positions within the company \cite{AlexyHenkel-87} and found that software testers and architects were particularly supportive of OSS involvement, indicating the potential for widespread employee. To navigate this cultural shift, companies can ``dually adapt'' to community and commercial habits \cite{mouakhar2017open}. This adaptation involves balancing community and commercial logic through strategic maneuvers. It is crucial to establish trust and foster collaboration by providing clear guidelines on intellectual property sharing and contribution processes. Investing in essential resources and partnerships can empower companies to contribute meaningfully while protecting their core interests.

\item[\textsc{IC2}] \textbf{Promoting companies' business capability.}

Intuitively, the greater a company's level of engagement, the more strategic its use of open source becomes. As a result, the company reaps higher value by participating in OSS projects. However, this requires increased resource investment, including human 
and financial costs \cite{Weiss-106, west2008designing}. Here comes a key issue that companies may face, i.e., higher coordination costs and risks to integrate internal and external innovation \cite{WestGallagher-29, ButlerGamalielsson-49, LindmanJuutilainen-173}, because it will inevitably bring changes in user experience, underlying code, and security. In addition, companies need to strike a balance, because the direction of OSS development is less likely to be in line with the company, and not all the features of OSS are welcomed by the company's customers. Enterprises have two coordination directions: change the development direction of the community or adjust their development direction. Link et al. studied the relationship between the open source phenomenon and the development of new enterprises \cite{LinkQureshiv-61}, and they found that the development of OSS benefits the establishment and development of new enterprises. From the reciprocity perspective, Yu \cite{yu2020role} explored how a firm’s open source strategy, characterized by its
participation in the OSS community, contributes to its financial performance. This study found that the OSS participation of companies taking technology from OSS directly affects their financial performance. On the contrary, the financial effects on companies that give technology to or socially participate (mutually exclusive to technical participation) in the OSS community, are fully mediated by the symbiotic relationship between the company and the open source community. 


However, some small companies are afraid to be the free labor to support the dominant company of the community when contributing to the OSS community \cite{yuxia2017openstack}. Nguyen et al. \cite{NguyenDucCruzes-60} examined how firms maintain both collaboration and competition and discovered that companies prioritize the co-creation of shared value and, to some extent, adapt in response to the competitive nature of open-source software ecosystems. Linåker et al. discovered that stakeholder influence and collaboration have developed over time as firms become more interconnected and interdependent \cite{linaaker2016firms}. Collaboration is essential for addressing complex challenges, even among companies with competing business models. Notably, firms recognize the value of collaboration and prioritize common goals over competition. 

\item[\textsc{IC3}] \textbf{Attract more companies and coordinate their behaviors.}
Beyond purely transactional benefits, Andersen-Gott et al. found a strong sense of moral obligation to contribute among surveyed companies \cite{Andersen-GottGhinea-27}. Despite this, the ``free-rider'' phenomenon, where companies benefit from the community without reciprocating, remains a concern \cite{yu2020role}. This way of OSS participation, driven by short-term gain, ultimately hinders long-term success. TanshoNoda et al. observed a significant gap between OSS utilization and contribution rates among companies \cite{TanshoNoda-97}. 
Feller et al. explored how to restrain companies in OSS communities \cite{FellerFinnegan-149}. They found that companies' behavior is restrained by their rights and interests toward the OSS community. If a company consistently engages in harmful practices in the community, it may be removed from the OSS network or its access may be restricted, thus affecting the company's reputation. In this way, companies are restrained from behaving themselves in the collaboration network.


As previously discussed in section \ref{tab: motivations}, technology acquisition is a key driver for company involvement. Interestingly, Butler et al. found that companies whose products are part of an OSS project tend to be more engaged, contributing to the community while simultaneously increasing their own product value \cite{ButlerGamalielsson-49}. This represents a perfect example of a mutually beneficial partnership. Commercial participants in OSS communities can attract deeper engagement from companies, leading to a thriving ecosystem of collaboration and innovation.
\end{enumerate}

\subsection{Impact on Developers}
Existing studies have explored the impact of commercial participation on both volunteers and paid developers. After categorizing, the identified impacts can be divided into two aspects: causing the loss of volunteers and suffering from role conflict.

\begin{enumerate}
    \item[\textsc{ID1}] \textbf{Cause the loss of volunteers.} The departure and retention of volunteers have a strong relation with the sustainability of OSS communities. Volunteers tend to contribute to OSS communities sponsored by non-profit organizations \cite{StewartGosain-155}. Because of the ``xenophobic'' characteristics of the community where companies are highly involved, it brings challenges for volunteers' long-term contribution. Osterloh et al. suggested that volunteers may cease contributing if the OSS community is dominated by commercial entities \cite{osterloh2002trust}. This is due to a lack of trust in the adherence of most contributors to the mutual norms of reciprocity within OSS communities. Although the increasing commercial participation in OSS is related to a decrease in the inflow of volunteers \cite{birkinbine2020incorporating}, not all kinds of company participation harm volunteer participation. Zhou et al. found that companies acting as collaborators in the OSS projects can increase volunteer participation \cite{zhou2016inflow}. Besides, the diversity of companies in an OSS community matters a lot in developer inflows. Zhang et al. measured the diversity of companies in OpenStack by the entropy of their contribution models and found a positive relationship between volunteer numbers and the diversity of companies \cite{zhang2019companies}. 


    In some company-sponsored OSS communities, the companies will set up several policies or measures to attract and retain contributors. Guizani et al., \cite{guizani2023rules} found that whether it is for internal employees or outside developers, companies hosting the OSS project usually try to encourage more contributors to participate. As for internal employees, the company may encourage employees to make contributions to OSS projects to obtain benefits. On the outside, the company will also use its influence to attract more contributors. For example, companies may organize some incentive activities, such as Google Summer of Code\footnote{https://summerofcode.withgoogle.com/} programs. These behaviors have effectively introduced more contributors to a certain extent and improved the sustainability of the OSS community. The acquisition of the hosting company will also have a significant impact on the OSS community \cite{agerfalkFitzgerald-180, JohriNov-95} because new and old parent companies may have different values and attitudes towards the OSS community. The most typical example is MySQL, which was taken over by Sun, and Sun was taken over by Oracle later. The evidence shows that the community's contribution increased after Sun's acquisition of MySQL, but decreased after Oracle's acquisition. This is partly because contributors believe that Sun is an OSS contributor, while Oracle is not.

    \item[\textsc{ID2}] \textbf{Role conflict for paid developers.} Paid developers play an important role in the OSS community with increasing commercial participation. On the one hand, paid developers need to fulfill the objectives of the companies that pay them to contribute to the community. On the other hand, they can be easily assimilated with the OSS community's open, transparent, and collaborative nature. In the survey conducted by Butler et al. \cite{ButlerGamalielsson-49}, paid developers are highly trusted by the organization, with one respondent even stating that about 90 percent of his work on the OSS project was not specifically requested or directed by the organization. Samuel et al. found that \cite{B.H.-120} paid developers have a clustering effect; the paid developers of the same company prefer to support each other and prefer to communicate internally. 
    According to Homscheid et al., \cite{HomscheidSchaarschmidt-100} the cultural conflict between company and community can affect the paid developers' sense of identity. Daniel et al. studied the influence of employees' perceptions of OSS ideology and other colleagues' perceptions of OSS on employee contribution \cite{DanielMaruping-56}, and they find that employees' commitment to their company reinforces the impact of their commitment to the OSS community in driving ongoing code contributions. Schaarschmidt et al. found that whether as part of the community or as part of the hiring organization, developers' imbalance in their understanding of ``self'' can lead to role conflict, and this kind of imbalance can improve turnover intentions \cite{SchaarschmidtStol-197}.  
    
\end{enumerate}

\subsection{Impact on Projects} \label{Impact On Project}
The impact of commercial participation on OSS projects has advantages and disadvantages: companies can enable OSS projects to flourish to some extent, but it can also pose risks to the long-term sustainability of the community \cite{Y.H.-30}.
In this section, we will discuss the investigated impact of commercial participation on OSS projects.

\begin{enumerate}
\item[\textsc{IP1}] \textbf{Affect project quality.} Commercial participation can help OSS projects become more popular and grow significantly with no need to solve the challenges of human and infrastructure resources \cite{newtonleveraging, CapraFrancalanci-78}. Meanwhile, the projects that are participated by firms tend to be larger and have more contributors than volunteers-powered ones \cite{A.D.-166}. Llanes et al. \cite{LlanesdeElejalde-23} found that commercial participation can provide OSS communities with higher-quality code and well-considered functionality. However, Capra et al. indicated that higher firms’ participation can lead to lower structural software design quality \cite{CapraFrancalanci-78}. 


\item[\textsc{IP2}] \textbf{Reduce project openness.} Commercial companies can play a major role in the decision-making and development of OSS projects \cite{CapiluppiBaravalle-94}. Closure represented by the company and openness represented by the community will lead to conflicts in the process of combination \cite{ONeilMuselli-50}. It can have an impact on intellectual property management \cite{A.D.-166}, as evidenced by the fact that the GNU general public license is scarce in commercially involved projects. Besides, to maintain benefits, companies may try to influence the direction of the community. Schaarschmidt et al. found that companies tend to rely on their resources or the resources of other companies to control a project \cite{SchaarschmidtVonKortzfleisch-65, SchaarschmidtWalsh-66}. As a consequence, the OSS community will reduce openness \cite{zhou2016inflow, west2008designing, Capra123, Y.H.-30}, thus threatening the sustainability \cite{ZhangStol-196}. For OSS projects at an early stage, commercial engagement is welcomed because of clear development direction and resources. In a mature community, it may become a disadvantage because an open, free, and fair governance model becomes more critical for the participation of volunteers and other organizations \cite{newtonleveraging, D.-150, zhang2019companies, ZhangStol-196}.

\end{enumerate}

\begin{summarybox}
    \textbf{Summary: } The relationship between commercial companies and OSS projects is a mutually beneficial endeavor. Within this dynamic, companies derive advantages from the openness of OSS projects, while OSS projects benefit from their contributions of code and developers. However, several negative impacts still exist for the companies themselves, for both paid and volunteer developers, and for the OSS projects. How to foster a harmonious relationship between companies and OSS communities will be a recurring topic in the future, and there is no answer yet.
    
    
\end{summarybox}

\section{Research Opportunities}\label{sec:opportunities} 

There is already extensive literature on commercial involvement in open source. Our study provides a snapshot of the research topics and the gained progress. In this section, we discuss some future research avenues that received scarce attention and are necessary for boosting the sustainability of company-involved OSS projects.


\textbf{Quantify the decline of user cost:} Although many articles qualitatively show that the cost for customers after commercial participation in open source is reduced \cite{Riehle-157, Fei-RongDan-128}, most of these studies are based on interviews or experience. there is a lack of articles to quantitatively explore how much cost advantage the company's customers can gain from it.


\textbf{Explore more OSS ecosystems:} Much of the current research is focused on a few specific, influential OSS communities, such as the Linux kernel, OpenStack, and the Apache community. It is well known that software developed in open source way covers a wide variety of categories, including operating systems, databases, deep learning, and applications. More research needs to be done on the countless OSS communities to compare the characteristics of companies' motivations, contribution models, and the brought impacts in the context of different OSS ecosystems with the consideration of project domains. By doing so, companies can have a clear cookbook about how to behave for a certain OSS project.

\textbf{Stand on one company's viewpoint to investigate all its involved OSS projects:} Existing literature mainly explores companies' participation in the context of a specific OSS project, few studies stand on one company's viewpoint \cite{capek2005history}. Hence, we can build an understanding of why and how companies contribute to the specific project. However, one company can have interests in multiple OSS projects and even open source own projects. With the natural constraint of time and resources, companies need to manage and balance their effort on all these involved OSS projects. Thus, focusing on single companies and exploring their overall OSS participation strategies can shed some light on how companies should embrace OSS. 

\textbf{A longitudinal study among diverse regions:} Most studies focus on global corporate participation as a whole. 
Researchers have explored how companies from specific regions participate in OSS, e.g., Hauge et al. studied Norwegian companies \cite{HaugeZiemer-168}, Tansho et al. studied Japanese companies \cite{TanshoNoda-97}, Lundell et al. studied OSS in Swedish \cite{LundellLings-89}, and Newton et al. studied country diversity in their study \cite{newtonleveraging}. From their results, we can see that there are differences in the participation of companies in different countries to some extent. There is no vertical research on the participation of companies in different countries in OSS projects. Studying the differences across countries can shed light on how cultural, economic, and even regulatory factors influence companies' engagement in collaborative development. It can help identify the unique challenges and opportunities faced by companies in different regions, enabling a more nuanced understanding of their motivations, strategies, and outcomes.
By uncovering these variations, researchers can identify best practices and lessons learned from specific countries that can be applied globally. This knowledge can inform policy decisions, industry initiatives, and educational programs aimed at fostering open source collaboration. Such knowledge can guide project managers, developers, and community leaders in making informed decisions and fostering inclusive and diverse collaborations.



\section{Limitations}\label{sec:limitations}
This section discusses the potential threats to the validity of this study. 

\textbf{Construct Limitations:} Our manual search strategy targeted three of the most relevant paper retrieval agencies for articles. 
Two iterations of the snowball sampling process expanded our search beyond these sites but were still limited by the references included in the selected articles. And because of the subjectivity of our researchers in manually reading and organizing articles, the wording of some articles may differ slightly from the original text. We developed detailed guidelines in the review protocol before the start of the review to ensure that the selection process was as unbiased as possible. As we screened papers, we documented the reasons for including or excluding them.

\textbf{Internal Limitations:} Since the research method, evaluation criteria, and research object vary between different articles, even for the same research area, and most of the studies in this review use non-standardized criteria to express the size of the effect, our article is to aggregate and synthesize all articles at the overall level. Some of the research analyzed in this paper does not focus on commercial participation in OSS communities or only mentions the relevant content. However, since commercial participation is a part of OSS communities, the articles that do not only focus on commercial participation can provide us with a more macro perspective on the phenomenon. Thus, we chose to include these articles in this study. 


\section{Related work}\label{sec: relatedwork}

There are several existing survey papers on commercial participation in OSS. In this section, we will analyze the current reviews and the differences with our work.

Wang et al. conducted a survey about motivations for individuals and companies to participate in open source projects \cite{Fei-RongDan-128}. They combined individuals and companies to describe motivations' function mechanism and divided motivations into three dimensions: i.e., economic, social, and technological. Similarly, we also follow the three dimensions to answer RQ2, the motivations of companies participating in OSS.
The study \cite{Fei-RongDan-128} strikes a balance between individuals and companies, which is different from ours.

Höst et al. conducted a systematic review of OSS in commercial software product development \cite{Host-77}. This work mainly studies how OSS can be deployed in commercial software. They only have a general description of the motivations and methods for companies to participate in OSS communities, which overlaps with our results of RQ2 and RQ3, but our research is broader and deeper.
Aberdour et al. reviewed the objective research on the quality of OSS, and business participation is only part of it \cite{aberdour2007achieving}. Van et al. introduced the development of commercial participation in open source in the form of introducing cases \cite{van2009commodification}. 

Osterloh et al. introduced open source, reviewed its history, and introduced the advantages of open source workflow \cite{OsterlohRota-164}. D., Riehle presented a series of reviews discussing open source business model \cite{Riehle-48, Riehle-157, Riehle-186}. Their works include many parts of OSS and commercial participation, which is a great inspiration for our work.  
Shahrokh Shahrivar et al. \cite{SHAHRIVAR2018202} provide a systematic review of commercial open source, which describes commercial open source business models and categorizes them into value proposition, value creation, and delivery, which have some overlaps with our RQ3's results, but our survey also examines the motivation, collaboration, and impact aspect of companies' participation in OSS.
 
\section{Conclusions}\label{sec: conclusions}

In this paper, we presented a comprehensive literature review about commercial participation in OSS. Specifically, we first collected and analyzed 92 research papers and we mainly divided our literature review into three themes: Motivations, Models, and Impacts. Then, we provide summary statistics for all selected papers based on full-text reading. We count the publication venue, research methods, and publication year, and create visualization graphs. We divided the motivation of commercial participation into three parts, economic, technical, and social, and discussed them separately. We discussed the model and paradigm for companies to participate in OSS projects and discussed the factors that affect the model. Fifth, we analyzed the impact of business participation and discussed the impact of business participation on ourselves, volunteers, open source projects, etc. 

For novice researchers and practitioners, our analysis of the commercial engagement in OSS projects can significantly reduce ambiguity and barriers to entry. In addition, the motivation, model, and implications of the summary provide a theoretical reference for (1) insightful ideas for future researchers to pursue promising research topics and solutions, and (2) future practitioners (companies or communities) to engage in business activities.

\bibliographystyle{ACM-Reference-Format}
\bibliography{references}


\begin{thebibliography}{122}


\ifx \showCODEN    \undefined \def \showCODEN     #1{\unskip}     \fi
\ifx \showDOI      \undefined \def \showDOI       #1{#1}\fi
\ifx \showISBNx    \undefined \def \showISBNx     #1{\unskip}     \fi
\ifx \showISBNxiii \undefined \def \showISBNxiii  #1{\unskip}     \fi
\ifx \showISSN     \undefined \def \showISSN      #1{\unskip}     \fi
\ifx \showLCCN     \undefined \def \showLCCN      #1{\unskip}     \fi
\ifx \shownote     \undefined \def \shownote      #1{#1}          \fi
\ifx \showarticletitle \undefined \def \showarticletitle #1{#1}   \fi
\ifx \showURL      \undefined \def \showURL       {\relax}        \fi
\providecommand\bibfield[2]{#2}
\providecommand\bibinfo[2]{#2}
\providecommand\natexlab[1]{#1}
\providecommand\showeprint[2][]{arXiv:#2}

\bibitem[Aberdour(2007)]%
        {aberdour2007achieving}
\bibfield{author}{\bibinfo{person}{Mark Aberdour}.} \bibinfo{year}{2007}\natexlab{}.
\newblock \showarticletitle{Achieving quality in open-source software}.
\newblock \bibinfo{journal}{\emph{IEEE software}} \bibinfo{volume}{24}, \bibinfo{number}{1} (\bibinfo{year}{2007}), \bibinfo{pages}{58--64}.
\newblock


\bibitem[ALEXY and HENKEL(2007)]%
        {AlexyHenkel-87}
\bibfield{author}{\bibinfo{person}{OLIVER ALEXY} {and} \bibinfo{person}{JOACHIM HENKEL}.} \bibinfo{year}{2007}\natexlab{}.
\newblock \showarticletitle{PROMOTING THE PENGUIN: WHO IS ADVOCATING OPEN SOURCE SOFTWARE IN COMMERCIAL SETTINGS?}
\newblock \bibinfo{journal}{\emph{Academy of Management Proceedings}} \bibinfo{volume}{2007}, \bibinfo{number}{1} (\bibinfo{year}{2007}), \bibinfo{pages}{1--6}.
\newblock
\urldef\tempurl%
\url{https://doi.org/10.5465/ambpp.2007.26530011}
\showDOI{\tempurl}
\newblock
\shownote{identifier: 10.5465/ambpp.2007.26530011}.


\bibitem[Andersen-Gott et~al\mbox{.}(2012)]%
        {Andersen-GottGhinea-27}
\bibfield{author}{\bibinfo{person}{Morten Andersen-Gott}, \bibinfo{person}{Gheorghita Ghinea}, {and} \bibinfo{person}{Bendik Bygstad}.} \bibinfo{year}{2012}\natexlab{}.
\newblock \showarticletitle{Why do commercial companies contribute to open source software?}
\newblock \bibinfo{journal}{\emph{International Journal of Information Management}} \bibinfo{volume}{32}, \bibinfo{number}{2} (\bibinfo{year}{2012}), \bibinfo{pages}{106--117}.
\newblock
\urldef\tempurl%
\url{https://doi.org/10.1016/j.ijinfomgt.2011.10.003}
\showDOI{\tempurl}


\bibitem[Baranes et~al\mbox{.}(2020)]%
        {BaranesVuong-54}
\bibfield{author}{\bibinfo{person}{E. Baranes}, \bibinfo{person}{C.~H. Vuong}, {and} \bibinfo{person}{M. Zeroukhi}.} \bibinfo{year}{2020}\natexlab{}.
\newblock \showarticletitle{Competitive strategy of proprietary software firms in an open source environment}.
\newblock \bibinfo{journal}{\emph{Review of Economic Research on Copyright Issues}} \bibinfo{volume}{17}, \bibinfo{number}{1} (\bibinfo{year}{2020}), \bibinfo{pages}{38--59}.
\newblock


\bibitem[Bauer et~al\mbox{.}(2020)]%
        {BauerHarutyunyan-136}
\bibfield{author}{\bibinfo{person}{Andreas Bauer}, \bibinfo{person}{Nikolay Harutyunyan}, \bibinfo{person}{Dirk Riehle}, {and} \bibinfo{person}{Georg-Daniel Schwarz}.} \bibinfo{year}{2020}\natexlab{}.
\newblock \showarticletitle{Challenges of Tracking and Documenting Open Source Dependencies in Products: A Case Study}.
\newblock \bibinfo{publisher}{Springer International Publishing}, \bibinfo{address}{Cham}, \bibinfo{pages}{25--35}.
\newblock


\bibitem[Biolchini et~al\mbox{.}(2005)]%
        {biolchini2005systematic}
\bibfield{author}{\bibinfo{person}{Jorge Biolchini}, \bibinfo{person}{Paula~Gomes Mian}, \bibinfo{person}{Ana Candida~Cruz Natali}, {and} \bibinfo{person}{Guilherme~Horta Travassos}.} \bibinfo{year}{2005}\natexlab{}.
\newblock \showarticletitle{Systematic review in software engineering}.
\newblock \bibinfo{journal}{\emph{System engineering and computer science department COPPE/UFRJ, Technical Report ES}} \bibinfo{volume}{679}, \bibinfo{number}{05} (\bibinfo{year}{2005}), \bibinfo{pages}{45}.
\newblock


\bibitem[Birkinbine(2020)]%
        {birkinbine2020incorporating}
\bibfield{author}{\bibinfo{person}{Benjamin Birkinbine}.} \bibinfo{year}{2020}\natexlab{}.
\newblock \bibinfo{booktitle}{\emph{Incorporating the digital commons: Corporate involvement in free and open source software}}.
\newblock \bibinfo{publisher}{University of Westminster Press}.
\newblock


\bibitem[Bonaccorsi et~al\mbox{.}(2007)]%
        {A.D.-166}
\bibfield{author}{\bibinfo{person}{Andrea Bonaccorsi}, \bibinfo{person}{Dario Lorenzi}, \bibinfo{person}{Monica Merito}, {and} \bibinfo{person}{Cristina Rossi}.} \bibinfo{year}{2007}\natexlab{}.
\newblock \showarticletitle{Business firms' engagement in community projects. Empirical evidence and further developments of the research}. In \bibinfo{booktitle}{\emph{First International Workshop on Emerging Trends in FLOSS Research and Development (FLOSS'07: ICSE Workshops 2007)}}. IEEE, \bibinfo{pages}{13--13}.
\newblock


\bibitem[Bonaccorsi and Rossi(2004)]%
        {BonaccorsiRossi-46}
\bibfield{author}{\bibinfo{person}{Andrea Bonaccorsi} {and} \bibinfo{person}{Cristina Rossi}.} \bibinfo{year}{2004}\natexlab{}.
\newblock \showarticletitle{Altruistic individuals, selfish firms? The structure of motivation in Open Source software}.
\newblock \bibinfo{journal}{\emph{First Monday}} \bibinfo{volume}{9}, \bibinfo{number}{1} (\bibinfo{year}{2004}).
\newblock
\urldef\tempurl%
\url{https://doi.org/10.5210/fm.v9i1.1113}
\showDOI{\tempurl}


\bibitem[Bonaccorsi and Rossi(2006)]%
        {bonaccorsi2006comparing}
\bibfield{author}{\bibinfo{person}{Andrea Bonaccorsi} {and} \bibinfo{person}{Cristina Rossi}.} \bibinfo{year}{2006}\natexlab{}.
\newblock \showarticletitle{Comparing motivations of individual programmers and firms to take part in the open source movement: From community to business}.
\newblock \bibinfo{journal}{\emph{Knowledge, Technology \& Policy}} \bibinfo{volume}{18}, \bibinfo{number}{4} (\bibinfo{year}{2006}), \bibinfo{pages}{40--64}.
\newblock


\bibitem[Butler et~al\mbox{.}(2019)]%
        {ButlerGamalielsson-49}
\bibfield{author}{\bibinfo{person}{Simon Butler}, \bibinfo{person}{Jonas Gamalielsson}, \bibinfo{person}{Bj{\"o}rn Lundell}, \bibinfo{person}{Christoffer Brax}, \bibinfo{person}{Johan Sj{\"o}berg}, \bibinfo{person}{Anders Mattsson}, \bibinfo{person}{Tomas Gustavsson}, \bibinfo{person}{Jonas Feist}, {and} \bibinfo{person}{Erik L{\"o}nroth}.} \bibinfo{year}{2019}\natexlab{}.
\newblock \showarticletitle{On company contributions to community open source software projects}.
\newblock \bibinfo{journal}{\emph{IEEE Transactions on Software Engineering}} \bibinfo{volume}{47}, \bibinfo{number}{7} (\bibinfo{year}{2019}), \bibinfo{pages}{1381--1401}.
\newblock


\bibitem[Butler et~al\mbox{.}(2018a)]%
        {butler2018investigation}
\bibfield{author}{\bibinfo{person}{Simon Butler}, \bibinfo{person}{Jonas Gamalielsson}, \bibinfo{person}{Bj{\"o}rn Lundell}, \bibinfo{person}{Per Jonsson}, \bibinfo{person}{Johan Sj{\"o}berg}, \bibinfo{person}{Anders Mattsson}, \bibinfo{person}{Niklas Rick{\"o}}, \bibinfo{person}{Tomas Gustavsson}, \bibinfo{person}{Jonas Feist}, \bibinfo{person}{Stefan Landemoo}, {et~al\mbox{.}}} \bibinfo{year}{2018}\natexlab{a}.
\newblock \showarticletitle{An investigation of work practices used by companies making contributions to established OSS projects}. In \bibinfo{booktitle}{\emph{Proceedings of the 40th International Conference on Software Engineering: Software Engineering in Practice}}. \bibinfo{pages}{201--210}.
\newblock


\bibitem[Butler et~al\mbox{.}(2018b)]%
        {S.J.-104}
\bibfield{author}{\bibinfo{person}{Simon Butler}, \bibinfo{person}{Jonas Gamalielsson}, \bibinfo{person}{Björn Lundell}, \bibinfo{person}{Per Jonsson}, \bibinfo{person}{Johan Sjöberg}, \bibinfo{person}{Anders Mattsson}, \bibinfo{person}{Niklas Rickö}, \bibinfo{person}{Tomas Gustavsson}, \bibinfo{person}{Jonas Feist}, \bibinfo{person}{Stefan Landemoo}, {et~al\mbox{.}}} \bibinfo{year}{2018}\natexlab{b}.
\newblock \showarticletitle{An investigation of work practices used by companies making contributions to established OSS projects}. In \bibinfo{booktitle}{\emph{Proceedings of the 40th International Conference on Software Engineering: Software Engineering in Practice}}. \bibinfo{pages}{201--210}.
\newblock


\bibitem[Capek et~al\mbox{.}(2005)]%
        {capek2005history}
\bibfield{author}{\bibinfo{person}{Peter~G Capek}, \bibinfo{person}{Steven~P Frank}, \bibinfo{person}{Steve Gerdt}, {and} \bibinfo{person}{David Shields}.} \bibinfo{year}{2005}\natexlab{}.
\newblock \showarticletitle{A history of IBM's open-source involvement and strategy}.
\newblock \bibinfo{journal}{\emph{IBM systems journal}} \bibinfo{volume}{44}, \bibinfo{number}{2} (\bibinfo{year}{2005}), \bibinfo{pages}{249--257}.
\newblock


\bibitem[Capiluppi et~al\mbox{.}(2010)]%
        {CapiluppiBaravalle-94}
\bibfield{author}{\bibinfo{person}{Andrea Capiluppi}, \bibinfo{person}{Andres Baravalle}, {and} \bibinfo{person}{Nick Heap}.} \bibinfo{year}{2010}\natexlab{}.
\newblock \showarticletitle{From "community" to "commercial" FLOSS: the case of Moodle}. In \bibinfo{booktitle}{\emph{Proceedings of the 3rd International Workshop on Emerging Trends in Free/Libre/Open Source Software Research and Development}} \emph{(\bibinfo{series}{Proceedings of the 3rd International Workshop on Emerging Trends in Free/Libre/Open Source Software Research and Development})}. \bibinfo{publisher}{ACM}, \bibinfo{address}{Cape Town, South Africa}, \bibinfo{pages}{11--16}.
\newblock


\bibitem[Capra et~al\mbox{.}(2008)]%
        {Capra123}
\bibfield{author}{\bibinfo{person}{Eugenio Capra}, \bibinfo{person}{Chiara Francalanci}, {and} \bibinfo{person}{Francesco Merlo}.} \bibinfo{year}{2008}\natexlab{}.
\newblock \showarticletitle{An empirical study on the relationship between software design quality, development effort and governance in open source projects}.
\newblock \bibinfo{journal}{\emph{IEEE Transactions on Software Engineering}} \bibinfo{volume}{34}, \bibinfo{number}{6} (\bibinfo{year}{2008}), \bibinfo{pages}{765--782}.
\newblock


\bibitem[Capra et~al\mbox{.}(2011)]%
        {CapraFrancalanci-78}
\bibfield{author}{\bibinfo{person}{Eugenio Capra}, \bibinfo{person}{Chiara Francalanci}, \bibinfo{person}{Francesco Merlo}, {and} \bibinfo{person}{Cristina Rossi-Lamastra}.} \bibinfo{year}{2011}\natexlab{}.
\newblock \showarticletitle{Firms’ involvement in Open Source projects: A trade-off between software structural quality and popularity}.
\newblock \bibinfo{journal}{\emph{Journal of Systems and Software}} \bibinfo{volume}{84}, \bibinfo{number}{1} (\bibinfo{year}{2011}), \bibinfo{pages}{144--161}.
\newblock
\urldef\tempurl%
\url{https://doi.org/10.1016/j.jss.2010.09.004}
\showDOI{\tempurl}


\bibitem[Colombo et~al\mbox{.}(2014)]%
        {ColomboPiva-72}
\bibfield{author}{\bibinfo{person}{Massimo~G. Colombo}, \bibinfo{person}{Evila Piva}, {and} \bibinfo{person}{Cristina Rossi-Lamastra}.} \bibinfo{year}{2014}\natexlab{}.
\newblock \showarticletitle{Open innovation and within-industry diversification in small and medium enterprises: The case of open source software firms}.
\newblock \bibinfo{journal}{\emph{Research Policy}} \bibinfo{volume}{43}, \bibinfo{number}{5} (\bibinfo{year}{2014}), \bibinfo{pages}{891--902}.
\newblock


\bibitem[Coughlan et~al\mbox{.}(2013)]%
        {CoughlanNoda-96}
\bibfield{author}{\bibinfo{person}{Shane Coughlan}, \bibinfo{person}{Tetsuo Noda}, {and} \bibinfo{person}{Terutaka Tansho}.} \bibinfo{year}{2013}\natexlab{}.
\newblock \showarticletitle{A case study of the collaborative approaches to sustain open source business models}. In \bibinfo{booktitle}{\emph{Proceedings of the 9th International Symposium on Open Collaboration}} \emph{(\bibinfo{series}{Proceedings of the 9th International Symposium on Open Collaboration})}. \bibinfo{publisher}{ACM}, \bibinfo{address}{Hong Kong, China}, \bibinfo{pages}{1--3}.
\newblock


\bibitem[D.(2021)]%
        {Riehle-48}
\bibfield{author}{\bibinfo{person}{Riehle D.}} \bibinfo{year}{2021}\natexlab{}.
\newblock \showarticletitle{The Open Source Distributor Business Model}.
\newblock \bibinfo{journal}{\emph{Computer}} \bibinfo{volume}{54}, \bibinfo{number}{12} (\bibinfo{year}{2021}), \bibinfo{pages}{99--103}.
\newblock
\urldef\tempurl%
\url{https://doi.org/10.1109/MC.2021.3112318}
\showDOI{\tempurl}
\newblock
\shownote{Computer Computer}.


\bibitem[Daffara(2007)]%
        {daffara2007business}
\bibfield{author}{\bibinfo{person}{Carlo Daffara}.} \bibinfo{year}{2007}\natexlab{}.
\newblock \showarticletitle{Business models in FLOSS-based companies}. In \bibinfo{booktitle}{\emph{Workshop presentatioon at the 3rd Conference on Open Source Systems (OSS 2007)}}.
\newblock


\bibitem[Dahlander and Magnusson(2008a)]%
        {dahlander2008firms}
\bibfield{author}{\bibinfo{person}{Linus Dahlander} {and} \bibinfo{person}{Mats Magnusson}.} \bibinfo{year}{2008}\natexlab{a}.
\newblock \showarticletitle{How do firms make use of open source communities?}
\newblock \bibinfo{journal}{\emph{Long range planning}} \bibinfo{volume}{41}, \bibinfo{number}{6} (\bibinfo{year}{2008}), \bibinfo{pages}{629--649}.
\newblock


\bibitem[Dahlander and Magnusson(2008b)]%
        {DahlanderMagnusson-42}
\bibfield{author}{\bibinfo{person}{Linus Dahlander} {and} \bibinfo{person}{Mats Magnusson}.} \bibinfo{year}{2008}\natexlab{b}.
\newblock \showarticletitle{How do Firms Make Use of Open Source Communities?}
\newblock \bibinfo{journal}{\emph{Long Range Planning}} \bibinfo{volume}{41}, \bibinfo{number}{6} (\bibinfo{year}{2008}), \bibinfo{pages}{629--649}.
\newblock


\bibitem[Dahlander and Magnusson(2005)]%
        {dahlander2005relationships}
\bibfield{author}{\bibinfo{person}{Linus Dahlander} {and} \bibinfo{person}{Mats~G Magnusson}.} \bibinfo{year}{2005}\natexlab{}.
\newblock \showarticletitle{Relationships between open source software companies and communities: Observations from Nordic firms}.
\newblock \bibinfo{journal}{\emph{Research policy}} \bibinfo{volume}{34}, \bibinfo{number}{4} (\bibinfo{year}{2005}), \bibinfo{pages}{481--493}.
\newblock


\bibitem[Daniel et~al\mbox{.}(2011)]%
        {DanielMaruping-187}
\bibfield{author}{\bibinfo{person}{Sherae Daniel}, \bibinfo{person}{Likoebe Maruping}, \bibinfo{person}{Marcelo Cataldo}, {and} \bibinfo{person}{James Herbsleb}.} \bibinfo{year}{2011}\natexlab{}.
\newblock \showarticletitle{When cultures clash: Participation in open source communities and its implications for organizational commitment}.
\newblock \bibinfo{journal}{\emph{ICIS}} (\bibinfo{year}{2011}).
\newblock


\bibitem[Daniel et~al\mbox{.}(2018)]%
        {DanielMaruping-56}
\bibfield{author}{\bibinfo{person}{Sherae~L. Daniel}, \bibinfo{person}{Likoebe~M. Maruping}, \bibinfo{person}{Marcelo Cataldo}, {and} \bibinfo{person}{Jin Herbsleb}.} \bibinfo{year}{2018}\natexlab{}.
\newblock \showarticletitle{The Impact of Ideology Misfit on Open Source Software Communities and Companies}.
\newblock \bibinfo{journal}{\emph{MIS quarterly}} \bibinfo{volume}{42}, \bibinfo{number}{4} (\bibinfo{year}{2018}), \bibinfo{pages}{1069}.
\newblock
\urldef\tempurl%
\url{https://doi.org/10.25300/MISQ/2018/14242}
\showDOI{\tempurl}


\bibitem[Dias et~al\mbox{.}(2018)]%
        {dias2018drives}
\bibfield{author}{\bibinfo{person}{Luis~Felipe Dias}, \bibinfo{person}{Igor Steinmacher}, {and} \bibinfo{person}{Gustavo Pinto}.} \bibinfo{year}{2018}\natexlab{}.
\newblock \showarticletitle{Who drives company-owned OSS projects: internal or external members?}
\newblock \bibinfo{journal}{\emph{Journal of the Brazilian Computer Society}} \bibinfo{volume}{24}, \bibinfo{number}{1} (\bibinfo{year}{2018}), \bibinfo{pages}{1--17}.
\newblock


\bibitem[Dijkers et~al\mbox{.}(2018)]%
        {J.R.-105}
\bibfield{author}{\bibinfo{person}{Joost Dijkers}, \bibinfo{person}{Rowan Sincic}, \bibinfo{person}{Nicole Wasankhasit}, {and} \bibinfo{person}{Slinger Jansen}.} \bibinfo{year}{2018}\natexlab{}.
\newblock \showarticletitle{Exploring the effect of software ecosystem health on the financial performance of the open source companies}. In \bibinfo{booktitle}{\emph{Proceedings of the 1st International Workshop on Software Health}}. \bibinfo{pages}{48--55}.
\newblock


\bibitem[Economides and Katsamakas(2006)]%
        {EconomidesKatsamakas-43}
\bibfield{author}{\bibinfo{person}{Nicholas Economides} {and} \bibinfo{person}{Evangelos Katsamakas}.} \bibinfo{year}{2006}\natexlab{}.
\newblock \showarticletitle{Two-Sided Competition of Proprietary vs. Open Source Technology Platforms and the Implications for the Software Industry}.
\newblock \bibinfo{journal}{\emph{Management Science}} \bibinfo{volume}{52}, \bibinfo{number}{7} (\bibinfo{year}{2006}), \bibinfo{pages}{1057--1071}.
\newblock
\urldef\tempurl%
\url{https://doi.org/10.1287/mnsc.1060.0549}
\showDOI{\tempurl}
\newblock
\shownote{identifier: 10.1287/mnsc.1060.0549}.


\bibitem[Ehls(2017)]%
        {ehls2017open}
\bibfield{author}{\bibinfo{person}{Daniel Ehls}.} \bibinfo{year}{2017}\natexlab{}.
\newblock \showarticletitle{Open source project collapse--sources and patterns of failure}.
\newblock  (\bibinfo{year}{2017}).
\newblock


\bibitem[Fei-Rong et~al\mbox{.}(2005)]%
        {Fei-RongDan-128}
\bibfield{author}{\bibinfo{person}{Wang Fei-Rong}, \bibinfo{person}{He Dan}, {and} \bibinfo{person}{Chen Jin}.} \bibinfo{year}{2005}\natexlab{}.
\newblock \showarticletitle{Motivations of Individuals and Firms Participating in Open Source Community}. In \bibinfo{booktitle}{\emph{2005 International Conference on Machine Learning and Cybernetics}} \emph{(\bibinfo{series}{2005 International Conference on Machine Learning and Cybernetics}, Vol.~\bibinfo{volume}{1})}. \bibinfo{publisher}{IEEE}, \bibinfo{pages}{309--314}.
\newblock


\bibitem[Feller et~al\mbox{.}(2008)]%
        {FellerFinnegan-149}
\bibfield{author}{\bibinfo{person}{Joseph Feller}, \bibinfo{person}{Patrick Finnegan}, \bibinfo{person}{Brian Fitzgerald}, {and} \bibinfo{person}{Jeremy Hayes}.} \bibinfo{year}{2008}\natexlab{}.
\newblock \showarticletitle{From peer production to productization: A study of socially enabled business exchanges in open source service networks}.
\newblock \bibinfo{journal}{\emph{Information Systems Research}} \bibinfo{volume}{19}, \bibinfo{number}{4} (\bibinfo{year}{2008}), \bibinfo{pages}{475--493}.
\newblock


\bibitem[Fendt et~al\mbox{.}(2016)]%
        {O.M.-152}
\bibfield{author}{\bibinfo{person}{Oliver Fendt}, \bibinfo{person}{Michael Jaeger}, {and} \bibinfo{person}{Ricardo~Jimenez Serrano}.} \bibinfo{year}{2016}\natexlab{}.
\newblock \showarticletitle{Industrial experience with open source software process management}. In \bibinfo{booktitle}{\emph{2016 IEEE 40th Annual Computer Software and Applications Conference (COMPSAC)}}, Vol.~\bibinfo{volume}{2}. IEEE, \bibinfo{pages}{180--185}.
\newblock


\bibitem[Fitzgerald(2006)]%
        {fitzgerald2006transformation}
\bibfield{author}{\bibinfo{person}{Brian Fitzgerald}.} \bibinfo{year}{2006}\natexlab{}.
\newblock \showarticletitle{The transformation of open source software}.
\newblock \bibinfo{journal}{\emph{MIS quarterly}} (\bibinfo{year}{2006}), \bibinfo{pages}{587--598}.
\newblock


\bibitem[Germonprez et~al\mbox{.}(2014)]%
        {GermonprezKendall-182}
\bibfield{author}{\bibinfo{person}{Matt Germonprez}, \bibinfo{person}{Julie~E. Kendall}, \bibinfo{person}{Kenneth~E. Kendall}, {and} \bibinfo{person}{Brett Young}.} \bibinfo{year}{2014}\natexlab{}.
\newblock \showarticletitle{Collectivism, creativity, competition, and control in open source software development: reflections on the emergent governance of the SPDX® working group}.
\newblock \bibinfo{journal}{\emph{International Journal of Information Systems and Management}} \bibinfo{volume}{1}, \bibinfo{number}{1-2} (\bibinfo{year}{2014}), \bibinfo{pages}{125--145}.
\newblock


\bibitem[Goggins et~al\mbox{.}(2021)]%
        {goggins2021making}
\bibfield{author}{\bibinfo{person}{Sean~P Goggins}, \bibinfo{person}{Matt Germonprez}, {and} \bibinfo{person}{Kevin Lumbard}.} \bibinfo{year}{2021}\natexlab{}.
\newblock \showarticletitle{Making open source project health transparent}.
\newblock \bibinfo{journal}{\emph{Computer}} \bibinfo{volume}{54}, \bibinfo{number}{8} (\bibinfo{year}{2021}), \bibinfo{pages}{104--111}.
\newblock


\bibitem[Gonzalez-Barahona et~al\mbox{.}(2013)]%
        {gonzalez2013understanding}
\bibfield{author}{\bibinfo{person}{Jesus~M Gonzalez-Barahona}, \bibinfo{person}{Daniel Izquierdo-Cortazar}, \bibinfo{person}{Stefano Maffulli}, {and} \bibinfo{person}{Gregorio Robles}.} \bibinfo{year}{2013}\natexlab{}.
\newblock \showarticletitle{Understanding how companies interact with free software communities}.
\newblock \bibinfo{journal}{\emph{IEEE software}} \bibinfo{volume}{30}, \bibinfo{number}{5} (\bibinfo{year}{2013}), \bibinfo{pages}{38--45}.
\newblock


\bibitem[Gonzalez-Barahona and Robles(2013)]%
        {gonzalez2013trends}
\bibfield{author}{\bibinfo{person}{Jesus~M Gonzalez-Barahona} {and} \bibinfo{person}{Gregorio Robles}.} \bibinfo{year}{2013}\natexlab{}.
\newblock \showarticletitle{Trends in free, libre, open source software communities: From volunteers to companies}.
\newblock \bibinfo{journal}{\emph{it--Information Technology it--Information Technology}} \bibinfo{volume}{55}, \bibinfo{number}{5} (\bibinfo{year}{2013}), \bibinfo{pages}{173--180}.
\newblock


\bibitem[Gruber and Henkel(2006)]%
        {gruber2006new}
\bibfield{author}{\bibinfo{person}{Marc Gruber} {and} \bibinfo{person}{Joachim Henkel}.} \bibinfo{year}{2006}\natexlab{}.
\newblock \showarticletitle{New ventures based on open innovation--an empirical analysis of start-up firms in embedded Linux}.
\newblock \bibinfo{journal}{\emph{International Journal of Technology Management}} \bibinfo{volume}{33}, \bibinfo{number}{4} (\bibinfo{year}{2006}), \bibinfo{pages}{356--372}.
\newblock


\bibitem[Guizani et~al\mbox{.}(2023)]%
        {guizani2023rules}
\bibfield{author}{\bibinfo{person}{Mariam Guizani}, \bibinfo{person}{Aileen~Abril Castro-Guzman}, \bibinfo{person}{Anita Sarma}, {and} \bibinfo{person}{Igor Steinmacher}.} \bibinfo{year}{2023}\natexlab{}.
\newblock \showarticletitle{Rules of Engagement: Why and How Companies Participate in OSS}.
\newblock \bibinfo{journal}{\emph{arXiv preprint arXiv:2303.08266}} (\bibinfo{year}{2023}).
\newblock


\bibitem[Guizani et~al\mbox{.}(2021)]%
        {GuizaniChatterjee-108}
\bibfield{author}{\bibinfo{person}{Mariam Guizani}, \bibinfo{person}{Amreeta Chatterjee}, \bibinfo{person}{Bianca Trinkenreich}, \bibinfo{person}{Mary~Evelyn May}, \bibinfo{person}{Geraldine~J. Noa-Guevara}, \bibinfo{person}{Liam~James Russell}, \bibinfo{person}{Griselda~G. Cuevas~Zambrano}, \bibinfo{person}{Daniel Izquierdo-Cortazar}, \bibinfo{person}{Igor Steinmacher}, \bibinfo{person}{Marco~A. Gerosa}, {and} \bibinfo{person}{Anita Sarma}.} \bibinfo{year}{2021}\natexlab{}.
\newblock \showarticletitle{The Long Road Ahead: Ongoing Challenges in Contributing to Large OSS Organizations and What to Do}.
\newblock \bibinfo{journal}{\emph{Proceedings of the ACM on Human-Computer Interaction}} \bibinfo{volume}{5}, \bibinfo{number}{CSCW2} (\bibinfo{year}{2021}), \bibinfo{pages}{1--30}.
\newblock
\urldef\tempurl%
\url{https://doi.org/10.1145/3479551}
\showDOI{\tempurl}
\newblock
\shownote{identifier: 10.1145/3479551}.


\bibitem[Gurbani et~al\mbox{.}(2006)]%
        {GurbaniGarvert-169}
\bibfield{author}{\bibinfo{person}{Vijay Gurbani}, \bibinfo{person}{Anita Garvert}, {and} \bibinfo{person}{James Herbsleb}.} \bibinfo{year}{2006}\natexlab{}.
\newblock \showarticletitle{A case study of a corporate open source development model}. In \bibinfo{booktitle}{\emph{Proceedings of the 28th international conference on Software engineering}} \emph{(\bibinfo{series}{Proceedings of the 28th international conference on Software engineering})}. \bibinfo{publisher}{ACM}, \bibinfo{address}{Shanghai, China}, \bibinfo{pages}{472--481}.
\newblock


\bibitem[Harutyunyan(2020)]%
        {harutyunyan2020managing}
\bibfield{author}{\bibinfo{person}{Nikolay Harutyunyan}.} \bibinfo{year}{2020}\natexlab{}.
\newblock \showarticletitle{Managing your open source supply chain-why and how?}
\newblock \bibinfo{journal}{\emph{Computer}} \bibinfo{volume}{53}, \bibinfo{number}{6} (\bibinfo{year}{2020}), \bibinfo{pages}{77--81}.
\newblock


\bibitem[Hauge et~al\mbox{.}(2007)]%
        {HaugeSorensen-167}
\bibfield{author}{\bibinfo{person}{O. Hauge}, \bibinfo{person}{C.~F. Sorensen}, {and} \bibinfo{person}{A. Rosdal}.} \bibinfo{year}{2007}\natexlab{}.
\newblock \bibinfo{booktitle}{\emph{Surveying industrial roles in open source software development}}. \bibinfo{series}{OPEN SOURCE DEVELOPMENT, ADOPTION AND INNOVATION}, Vol.~\bibinfo{volume}{234}.
\newblock \bibinfo{publisher}{Springer}.
\newblock


\bibitem[Hauge and Ziemer(2009)]%
        {HaugeZiemer-168}
\bibfield{author}{\bibinfo{person}{O. Hauge} {and} \bibinfo{person}{S. Ziemer}.} \bibinfo{year}{2009}\natexlab{}.
\newblock \bibinfo{booktitle}{\emph{Providing Commercial Open Source Software: Lessons Learned}}. \bibinfo{series}{OPEN SOURCE ECOSYSTEMS-DIVERSE COMMUNITIES INTERACTING}, Vol.~\bibinfo{volume}{299}.
\newblock \bibinfo{publisher}{Springer}.
\newblock


\bibitem[Hippel and Krogh(2003)]%
        {hippel2003open}
\bibfield{author}{\bibinfo{person}{Eric~von Hippel} {and} \bibinfo{person}{Georg~von Krogh}.} \bibinfo{year}{2003}\natexlab{}.
\newblock \showarticletitle{Open source software and the “private-collective” innovation model: Issues for organization science}.
\newblock \bibinfo{journal}{\emph{Organization science}} \bibinfo{volume}{14}, \bibinfo{number}{2} (\bibinfo{year}{2003}), \bibinfo{pages}{209--223}.
\newblock


\bibitem[Ho and Rai(2017)]%
        {HoRai-58}
\bibfield{author}{\bibinfo{person}{Shuk~Ying Ho} {and} \bibinfo{person}{Arun Rai}.} \bibinfo{year}{2017}\natexlab{}.
\newblock \showarticletitle{Continued Voluntary Participation Intention in Firm-Participating Open Source Software Projects}.
\newblock \bibinfo{journal}{\emph{Information Systems Research}} \bibinfo{volume}{28}, \bibinfo{number}{3} (\bibinfo{year}{2017}), \bibinfo{pages}{603--625}.
\newblock
\urldef\tempurl%
\url{https://doi.org/10.1287/isre.2016.0687}
\showDOI{\tempurl}
\newblock
\shownote{identifier: 10.1287/isre.2016.0687}.


\bibitem[Homscheid and Schaarschmidt(2016)]%
        {HomscheidSchaarschmidt-100}
\bibfield{author}{\bibinfo{person}{Dirk Homscheid} {and} \bibinfo{person}{Mario Schaarschmidt}.} \bibinfo{year}{2016}\natexlab{}.
\newblock \showarticletitle{Between organization and community: investigating turnover intention factors of firm-sponsored open source software developers}. In \bibinfo{booktitle}{\emph{Proceedings of the 8th ACM Conference on Web Science}} \emph{(\bibinfo{series}{Proceedings of the 8th ACM Conference on Web Science})}. \bibinfo{publisher}{ACM}, \bibinfo{address}{Hannover, Germany}, \bibinfo{pages}{336--337}.
\newblock


\bibitem[Höst and Oručević-Alagić(2011)]%
        {Host-77}
\bibfield{author}{\bibinfo{person}{Martin Höst} {and} \bibinfo{person}{Alma Oručević-Alagić}.} \bibinfo{year}{2011}\natexlab{}.
\newblock \showarticletitle{A systematic review of research on open source software in commercial software product development}.
\newblock \bibinfo{journal}{\emph{Information and Software Technology}} \bibinfo{volume}{53}, \bibinfo{number}{6} (\bibinfo{year}{2011}), \bibinfo{pages}{616--624}.
\newblock
\urldef\tempurl%
\url{https://doi.org/10.1016/j.infsof.2010.12.009}
\showDOI{\tempurl}


\bibitem[J. and S.(2005)]%
        {J.S.-31}
\bibfield{author}{\bibinfo{person}{West J.} {and} \bibinfo{person}{O'Mahony S.}} \bibinfo{year}{2005}\natexlab{}.
\newblock \showarticletitle{Contrasting Community Building in Sponsored and Community Founded Open Source Projects}. In \bibinfo{booktitle}{\emph{Proceedings of the 38th Annual Hawaii International Conference on System Sciences. IEEE, 2005.}} \emph{(\bibinfo{series}{Proceedings of the 38th Annual Hawaii International Conference on System Sciences. IEEE, 2005.})}. \bibinfo{publisher}{IEEE}, \bibinfo{pages}{196c--196c}.
\newblock


\bibitem[Johri et~al\mbox{.}(2011)]%
        {JohriNov-95}
\bibfield{author}{\bibinfo{person}{Aditya Johri}, \bibinfo{person}{Oded Nov}, {and} \bibinfo{person}{Raktim Mitra}.} \bibinfo{year}{2011}\natexlab{}.
\newblock \showarticletitle{"Cool" or "monster"?: company takeovers and their effect on open source community participation}. In \bibinfo{booktitle}{\emph{Proceedings of the 2011 iConference}} \emph{(\bibinfo{series}{Proceedings of the 2011 iConference})}. \bibinfo{publisher}{ACM}, \bibinfo{address}{Seattle, Washington, USA}, \bibinfo{pages}{327--331}.
\newblock


\bibitem[Kitchenham and Charters(2007)]%
        {2007Guidelines}
\bibfield{author}{\bibinfo{person}{B.~A. Kitchenham} {and} \bibinfo{person}{S. Charters}.} \bibinfo{year}{2007}\natexlab{}.
\newblock \showarticletitle{Guidelines for performing Systematic Literature Reviews in Software Engineering}.
\newblock  (\bibinfo{year}{2007}).
\newblock


\bibitem[Laurent(2004)]%
        {laurent2004understanding}
\bibfield{author}{\bibinfo{person}{Andrew M~St Laurent}.} \bibinfo{year}{2004}\natexlab{}.
\newblock \bibinfo{booktitle}{\emph{Understanding open source and free software licensing: guide to navigating licensing issues in existing \& new software}}.
\newblock \bibinfo{publisher}{" O'Reilly Media, Inc."}.
\newblock


\bibitem[Lerner and Tirole(2001)]%
        {lerner2001open}
\bibfield{author}{\bibinfo{person}{Josh Lerner} {and} \bibinfo{person}{Jean Tirole}.} \bibinfo{year}{2001}\natexlab{}.
\newblock \showarticletitle{The open source movement: Key research questions}.
\newblock \bibinfo{journal}{\emph{European economic review}} \bibinfo{volume}{45}, \bibinfo{number}{4-6} (\bibinfo{year}{2001}), \bibinfo{pages}{819--826}.
\newblock


\bibitem[Letellier(2008)]%
        {letellier2008open}
\bibfield{author}{\bibinfo{person}{Fran{\c{c}}ois Letellier}.} \bibinfo{year}{2008}\natexlab{}.
\newblock \showarticletitle{Open source software: the role of nonprofits in federating business and innovation ecosystems}.
\newblock \bibinfo{journal}{\emph{AFTE 2008}} (\bibinfo{year}{2008}).
\newblock


\bibitem[Lin{\aa}ker et~al\mbox{.}(2016)]%
        {linaaker2016firms}
\bibfield{author}{\bibinfo{person}{Johan Lin{\aa}ker}, \bibinfo{person}{Patrick Rempel}, \bibinfo{person}{Bj{\"o}rn Regnell}, {and} \bibinfo{person}{Patrick M{\"a}der}.} \bibinfo{year}{2016}\natexlab{}.
\newblock \showarticletitle{How firms adapt and interact in open source ecosystems: analyzing stakeholder influence and collaboration patterns}. In \bibinfo{booktitle}{\emph{International Working Conference on Requirements Engineering: Foundation for Software Quality}}. Springer, \bibinfo{pages}{63--81}.
\newblock


\bibitem[Lindman et~al\mbox{.}(2009)]%
        {LindmanJuutilainen-173}
\bibfield{author}{\bibinfo{person}{J. Lindman}, \bibinfo{person}{J.~P. Juutilainen}, {and} \bibinfo{person}{M. Rossi}.} \bibinfo{year}{2009}\natexlab{}.
\newblock \bibinfo{booktitle}{\emph{Beyond the Business Model: Incentives for Organizations to Publish Software Source Code}}. \bibinfo{series}{OPEN SOURCE ECOSYSTEMS-DIVERSE COMMUNITIES INTERACTING}, Vol.~\bibinfo{volume}{299}.
\newblock \bibinfo{publisher}{Springer Berlin Heidelberg}, \bibinfo{address}{Berlin, Heidelberg}.
\newblock


\bibitem[Link and Jeske(2017)]%
        {link2017understanding}
\bibfield{author}{\bibinfo{person}{Georg~JP Link} {and} \bibinfo{person}{Debora Jeske}.} \bibinfo{year}{2017}\natexlab{}.
\newblock \showarticletitle{Understanding organization and open source community relations through the attraction-selection-attrition model}. In \bibinfo{booktitle}{\emph{Proceedings of the 13th International Symposium on Open Collaboration}}. \bibinfo{pages}{1--8}.
\newblock


\bibitem[Link and Qureshi(2017)]%
        {LinkQureshiv-61}
\bibfield{author}{\bibinfo{person}{G.~J.~P. Link} {and} \bibinfo{person}{S. Qureshi}.} \bibinfo{year}{2017}\natexlab{}.
\newblock \showarticletitle{The role of open source in new business formation: Innovations for development}, Vol.~\bibinfo{volume}{2017-August}. \bibinfo{publisher}{Americas Conference on Information Systems}.
\newblock


\bibitem[Llanes and de~Elejalde(2013)]%
        {LlanesdeElejalde-23}
\bibfield{author}{\bibinfo{person}{Gastón Llanes} {and} \bibinfo{person}{Ramiro de Elejalde}.} \bibinfo{year}{2013}\natexlab{}.
\newblock \showarticletitle{Industry equilibrium with open-source and proprietary firms}.
\newblock \bibinfo{journal}{\emph{International Journal of Industrial Organization}} \bibinfo{volume}{31}, \bibinfo{number}{1} (\bibinfo{year}{2013}), \bibinfo{pages}{36--49}.
\newblock


\bibitem[Lundell et~al\mbox{.}(2006)]%
        {LundellLings-89}
\bibfield{author}{\bibinfo{person}{Björn Lundell}, \bibinfo{person}{Brian Lings}, {and} \bibinfo{person}{Edvin Lindqvist}.} \bibinfo{year}{2006}\natexlab{}.
\newblock \showarticletitle{Perceptions and Uptake of Open Source in Swedish Organisations}.
\newblock In \bibinfo{booktitle}{\emph{IFIP International Federation for Information Processing}}, \bibfield{editor}{\bibinfo{person}{E.~Damiani}, \bibinfo{person}{B.~Fitzgerald}, \bibinfo{person}{W.~Scacchi}, \bibinfo{person}{M.~Scotto}, {and} \bibinfo{person}{G.~Succi}} (Eds.). \bibinfo{series}{IFIP International Federation for Information Processing}, Vol.~\bibinfo{volume}{203}. \bibinfo{publisher}{Springer US}, \bibinfo{address}{Boston, MA}, \bibinfo{pages}{155--163}.
\newblock


\bibitem[Lundell et~al\mbox{.}(2010)]%
        {lundell2010open}
\bibfield{author}{\bibinfo{person}{Bj{\"o}rn Lundell}, \bibinfo{person}{Brian Lings}, {and} \bibinfo{person}{Edvin Lindqvist}.} \bibinfo{year}{2010}\natexlab{}.
\newblock \showarticletitle{Open source in Swedish companies: where are we?}
\newblock \bibinfo{journal}{\emph{Information Systems Journal}} \bibinfo{volume}{20}, \bibinfo{number}{6} (\bibinfo{year}{2010}), \bibinfo{pages}{519--535}.
\newblock


\bibitem[M.(2007)]%
        {M.-28}
\bibfield{author}{\bibinfo{person}{Aberdour M.}} \bibinfo{year}{2007}\natexlab{}.
\newblock \showarticletitle{Achieving Quality in Open-Source Software}.
\newblock \bibinfo{journal}{\emph{IEEE Software}} \bibinfo{volume}{24}, \bibinfo{number}{1} (\bibinfo{year}{2007}), \bibinfo{pages}{58--64}.
\newblock
\urldef\tempurl%
\url{https://doi.org/10.1109/MS.2007.2}
\showDOI{\tempurl}


\bibitem[Marsan et~al\mbox{.}(2012)]%
        {marsan2012adoption}
\bibfield{author}{\bibinfo{person}{Josianne Marsan}, \bibinfo{person}{Guy Par{\'e}}, {and} \bibinfo{person}{Anne Beaudry}.} \bibinfo{year}{2012}\natexlab{}.
\newblock \showarticletitle{Adoption of open source software in organizations: A socio-cognitive perspective}.
\newblock \bibinfo{journal}{\emph{The Journal of Strategic Information Systems}} \bibinfo{volume}{21}, \bibinfo{number}{4} (\bibinfo{year}{2012}), \bibinfo{pages}{257--273}.
\newblock


\bibitem[Morgan and Finnegan(2014)]%
        {morgan2014beyond}
\bibfield{author}{\bibinfo{person}{Lorraine Morgan} {and} \bibinfo{person}{Patrick Finnegan}.} \bibinfo{year}{2014}\natexlab{}.
\newblock \showarticletitle{Beyond free software: An exploration of the business value of strategic open source}.
\newblock \bibinfo{journal}{\emph{The Journal of Strategic Information Systems}} \bibinfo{volume}{23}, \bibinfo{number}{3} (\bibinfo{year}{2014}), \bibinfo{pages}{226--238}.
\newblock


\bibitem[Mouakhar and Tellier(2017)]%
        {mouakhar2017open}
\bibfield{author}{\bibinfo{person}{Khaireddine Mouakhar} {and} \bibinfo{person}{Alb{\'e}ric Tellier}.} \bibinfo{year}{2017}\natexlab{}.
\newblock \showarticletitle{How do Open Source software companies respond to institutional pressures? A business model perspective}.
\newblock \bibinfo{journal}{\emph{Journal of Enterprise Information Management}} \bibinfo{volume}{30}, \bibinfo{number}{4} (\bibinfo{year}{2017}), \bibinfo{pages}{534--554}.
\newblock


\bibitem[Munga et~al\mbox{.}(2009)]%
        {munga2009adoption}
\bibfield{author}{\bibinfo{person}{Neeshal Munga}, \bibinfo{person}{Thomas Fogwill}, {and} \bibinfo{person}{Quentin Williams}.} \bibinfo{year}{2009}\natexlab{}.
\newblock \showarticletitle{The adoption of open source software in business models: a Red Hat and IBM case study}. In \bibinfo{booktitle}{\emph{Proceedings of the 2009 Annual Research Conference of the South African Institute of Computer Scientists and Information Technologists}}. ACM, \bibinfo{pages}{112--121}.
\newblock


\bibitem[Newton and Fiore(2023)]%
        {newtonleveraging}
\bibfield{author}{\bibinfo{person}{Olivia~B Newton} {and} \bibinfo{person}{Stephen~M Fiore}.} \bibinfo{year}{2023}\natexlab{}.
\newblock \showarticletitle{Leveraging Corporate Engagement for Diversity in Free/Libre and Open Source Software Projects}.
\newblock  (\bibinfo{year}{2023}).
\newblock


\bibitem[Nguyen~Duc et~al\mbox{.}(2011)]%
        {nguyen2011impact}
\bibfield{author}{\bibinfo{person}{Anh Nguyen~Duc}, \bibinfo{person}{Daniela~S Cruzes}, \bibinfo{person}{Claudia Ayala}, {and} \bibinfo{person}{Reidar Conradi}.} \bibinfo{year}{2011}\natexlab{}.
\newblock \showarticletitle{Impact of stakeholder type and collaboration on issue resolution time in oss projects}. In \bibinfo{booktitle}{\emph{IFIP International conference on open source systems}}. Springer, \bibinfo{pages}{1--16}.
\newblock


\bibitem[Nguyen~Duc et~al\mbox{.}(2017)]%
        {NguyenDucCruzes-60}
\bibfield{author}{\bibinfo{person}{Anh Nguyen~Duc}, \bibinfo{person}{Daniela~S. Cruzes}, \bibinfo{person}{Geir~K. Hanssen}, \bibinfo{person}{Terje Snarby}, {and} \bibinfo{person}{Pekka Abrahamsson}.} \bibinfo{year}{2017}\natexlab{}.
\newblock \showarticletitle{Coopetition of Software Firms in Open Source Software Ecosystems}.
\newblock Vol.~\bibinfo{volume}{304}. \bibinfo{publisher}{Springer International Publishing}, \bibinfo{address}{Cham}, \bibinfo{pages}{146--160}.
\newblock


\bibitem[O~Neil et~al\mbox{.}(2021)]%
        {ONeilMuselli-50}
\bibfield{author}{\bibinfo{person}{Mathieu O~Neil}, \bibinfo{person}{Laure Muselli}, \bibinfo{person}{Mahin Raissi}, {and} \bibinfo{person}{Stefano Zacchiroli}.} \bibinfo{year}{2021}\natexlab{}.
\newblock \showarticletitle{‘Open source has won and lost the war’: Legitimising commercial–communal hybridisation in a FOSS project}.
\newblock \bibinfo{journal}{\emph{New Media \& Society}} \bibinfo{volume}{23}, \bibinfo{number}{5} (\bibinfo{year}{2021}), \bibinfo{pages}{1157--1180}.
\newblock
\urldef\tempurl%
\url{https://doi.org/10.1177/1461444820907022}
\showDOI{\tempurl}


\bibitem[Osterloh and Rota(2007)]%
        {OsterlohRota-164}
\bibfield{author}{\bibinfo{person}{Margit Osterloh} {and} \bibinfo{person}{Sandra Rota}.} \bibinfo{year}{2007}\natexlab{}.
\newblock \showarticletitle{Open source software development—Just another case of collective invention?}
\newblock \bibinfo{journal}{\emph{Research Policy}} \bibinfo{volume}{36}, \bibinfo{number}{2} (\bibinfo{year}{2007}), \bibinfo{pages}{157--171}.
\newblock
\urldef\tempurl%
\url{https://doi.org/10.1016/j.respol.2006.10.004}
\showDOI{\tempurl}


\bibitem[Osterloh et~al\mbox{.}(2002)]%
        {osterloh2002trust}
\bibfield{author}{\bibinfo{person}{Margit Osterloh}, \bibinfo{person}{Sandra Rota}, {and} \bibinfo{person}{Bernhard Kuster}.} \bibinfo{year}{2002}\natexlab{}.
\newblock \bibinfo{booktitle}{\emph{Trust and commerce in open source--a contradiction?}}
\newblock \bibinfo{publisher}{na}.
\newblock


\bibitem[O’Mahony(2007)]%
        {o2007governance}
\bibfield{author}{\bibinfo{person}{Siobh{\'a}n O’Mahony}.} \bibinfo{year}{2007}\natexlab{}.
\newblock \showarticletitle{The governance of open source initiatives: what does it mean to be community managed?}
\newblock \bibinfo{journal}{\emph{Journal of Management \& Governance}}  \bibinfo{volume}{11} (\bibinfo{year}{2007}), \bibinfo{pages}{139--150}.
\newblock


\bibitem[O’Neil et~al\mbox{.}(2021)]%
        {o2021coproduction}
\bibfield{author}{\bibinfo{person}{Mathieu O’Neil}, \bibinfo{person}{Xiaolan Cai}, \bibinfo{person}{Laure Muselli}, \bibinfo{person}{Fred Pailler}, {and} \bibinfo{person}{Stefano Zacchiroli}.} \bibinfo{year}{2021}\natexlab{}.
\newblock \bibinfo{booktitle}{\emph{The coproduction of open source software by volunteers and big tech firms}}.
\newblock \bibinfo{publisher}{News and Media Research Centre}.
\newblock


\bibitem[Raymond(2001)]%
        {2001The}
\bibfield{author}{\bibinfo{person}{E.~S. Raymond}.} \bibinfo{year}{2001}\natexlab{}.
\newblock \bibinfo{booktitle}{\emph{The Cathedral and the Bazaar: Musings on Linux and Open Source by an Accidental Revolutionary}}.
\newblock \bibinfo{publisher}{The cathedral and the bazaar - musings on Linux and open source by an accidental revoltionary (rev. ed.)}.
\newblock


\bibitem[Riehle(2007)]%
        {D.-150}
\bibfield{author}{\bibinfo{person}{Dirk Riehle}.} \bibinfo{year}{2007}\natexlab{}.
\newblock \showarticletitle{The economic motivation of open source software: Stakeholder perspectives}.
\newblock \bibinfo{journal}{\emph{Computer}} \bibinfo{volume}{40}, \bibinfo{number}{4} (\bibinfo{year}{2007}), \bibinfo{pages}{25--32}.
\newblock


\bibitem[Riehle(2009)]%
        {Riehle-153}
\bibfield{author}{\bibinfo{person}{D. Riehle}.} \bibinfo{year}{2009}\natexlab{}.
\newblock \bibinfo{booktitle}{\emph{The Commercial Open Source Business Model}}. \bibinfo{series}{VALUE CREATION IN E-BUSINESS MANAGEMENT}, Vol.~\bibinfo{volume}{36}.
\newblock \bibinfo{publisher}{Springer}.
\newblock


\bibitem[Riehle(2012)]%
        {Riehle-157}
\bibfield{author}{\bibinfo{person}{Dirk Riehle}.} \bibinfo{year}{2012}\natexlab{}.
\newblock \showarticletitle{The single-vendor commercial open course business model}.
\newblock \bibinfo{journal}{\emph{Information Systems and e-Business Management}}  \bibinfo{volume}{10} (\bibinfo{year}{2012}), \bibinfo{pages}{5--17}.
\newblock


\bibitem[Riehle(2019)]%
        {Riehle-186}
\bibfield{author}{\bibinfo{person}{Dirk Riehle}.} \bibinfo{year}{2019}\natexlab{}.
\newblock \showarticletitle{The innovations of open source}.
\newblock \bibinfo{journal}{\emph{Computer}} \bibinfo{volume}{52}, \bibinfo{number}{4} (\bibinfo{year}{2019}), \bibinfo{pages}{59--63}.
\newblock


\bibitem[Riehle(2020)]%
        {riehle2020single}
\bibfield{author}{\bibinfo{person}{Dirk Riehle}.} \bibinfo{year}{2020}\natexlab{}.
\newblock \showarticletitle{Single-vendor open source firms}.
\newblock \bibinfo{journal}{\emph{Computer}} \bibinfo{volume}{53}, \bibinfo{number}{4} (\bibinfo{year}{2020}), \bibinfo{pages}{68--72}.
\newblock


\bibitem[Robles et~al\mbox{.}(2007)]%
        {RoblesDuenas-177}
\bibfield{author}{\bibinfo{person}{G. Robles}, \bibinfo{person}{S. Duenas}, {and} \bibinfo{person}{J.~M. Gonzalez-Barahona}.} \bibinfo{year}{2007}\natexlab{}.
\newblock \bibinfo{booktitle}{\emph{Corporate involvement of libre software: Study of presence in Debian code over time}}. \bibinfo{series}{OPEN SOURCE DEVELOPMENT, ADOPTION AND INNOVATION}, Vol.~\bibinfo{volume}{234}.
\newblock \bibinfo{publisher}{Springer US}, \bibinfo{address}{Boston, MA}.
\newblock
\newblock
\shownote{Open Source Development, Adoption and Innovation}.


\bibitem[S. et~al\mbox{.}(2006)]%
        {S.G.-127}
\bibfield{author}{\bibinfo{person}{Valverde S.}, \bibinfo{person}{Theraulaz G.}, \bibinfo{person}{Gautrais J.}, \bibinfo{person}{Fourcassie V.}, {and} \bibinfo{person}{V.~Sole R.}} \bibinfo{year}{2006}\natexlab{}.
\newblock \showarticletitle{Self-organization patterns in wasp and open source communities}.
\newblock \bibinfo{journal}{\emph{IEEE Intelligent Systems}} \bibinfo{volume}{21}, \bibinfo{number}{2} (\bibinfo{year}{2006}), \bibinfo{pages}{36--40}.
\newblock
\urldef\tempurl%
\url{https://doi.org/10.1109/MIS.2006.34}
\showDOI{\tempurl}


\bibitem[Samuel et~al\mbox{.}(2021)]%
        {B.H.-120}
\bibfield{author}{\bibinfo{person}{Binny~M Samuel}, \bibinfo{person}{Hillol Bala}, \bibinfo{person}{Sherae~L Daniel}, {and} \bibinfo{person}{V Ramesh}.} \bibinfo{year}{2021}\natexlab{}.
\newblock \showarticletitle{Deconstructing the Nature of Collaboration in Organizations Open Source Software Development: The Impact of Developer and Task Characteristics}.
\newblock \bibinfo{journal}{\emph{IEEE Transactions on Software Engineering}} \bibinfo{volume}{48}, \bibinfo{number}{10} (\bibinfo{year}{2021}), \bibinfo{pages}{3969--3987}.
\newblock


\bibitem[Schaarschmidt and Stol(2018)]%
        {SchaarschmidtStol-197}
\bibfield{author}{\bibinfo{person}{Mario Schaarschmidt} {and} \bibinfo{person}{Klaas-Jan Stol}.} \bibinfo{year}{2018}\natexlab{}.
\newblock \showarticletitle{Company soldiers and gone-natives: role conflict and career ambition among firm-employed open source developers}.
\newblock  (\bibinfo{year}{2018}).
\newblock


\bibitem[Schaarschmidt and Von~Kortzfleisch(2015)]%
        {SchaarschmidtVonKortzfleisch-65}
\bibfield{author}{\bibinfo{person}{Mario Schaarschmidt} {and} \bibinfo{person}{Harald Von~Kortzfleisch}.} \bibinfo{year}{2015}\natexlab{}.
\newblock \showarticletitle{Firms' Resource Deployment and Project Leadership in Open Source Software Development}.
\newblock \bibinfo{journal}{\emph{International Journal of Innovation and Technology Management}} \bibinfo{volume}{12}, \bibinfo{number}{02} (\bibinfo{year}{2015}), \bibinfo{pages}{1550010}.
\newblock
\urldef\tempurl%
\url{https://doi.org/10.1142/S0219877015500108}
\showDOI{\tempurl}
\newblock
\shownote{identifier: 10.1142/S0219877015500108}.


\bibitem[Schaarschmidt et~al\mbox{.}(2015)]%
        {SchaarschmidtWalsh-66}
\bibfield{author}{\bibinfo{person}{Mario Schaarschmidt}, \bibinfo{person}{Gianfranco Walsh}, {and} \bibinfo{person}{Harald F.~O. von Kortzfleisch}.} \bibinfo{year}{2015}\natexlab{}.
\newblock \showarticletitle{How do firms influence open source software communities? A framework and empirical analysis of different governance modes}.
\newblock \bibinfo{journal}{\emph{Information and Organization}} \bibinfo{volume}{25}, \bibinfo{number}{2} (\bibinfo{year}{2015}), \bibinfo{pages}{99--114}.
\newblock
\urldef\tempurl%
\url{https://doi.org/10.1016/j.infoandorg.2015.03.001}
\showDOI{\tempurl}


\bibitem[Sen et~al\mbox{.}(2012)]%
        {sen2012open}
\bibfield{author}{\bibinfo{person}{Ravi Sen}, \bibinfo{person}{Siddhartha~S Singh}, {and} \bibinfo{person}{Sharad Borle}.} \bibinfo{year}{2012}\natexlab{}.
\newblock \showarticletitle{Open source software success: Measures and analysis}.
\newblock \bibinfo{journal}{\emph{Decision Support Systems}} \bibinfo{volume}{52}, \bibinfo{number}{2} (\bibinfo{year}{2012}), \bibinfo{pages}{364--372}.
\newblock


\bibitem[Shahrivar et~al\mbox{.}(2018)]%
        {SHAHRIVAR2018202}
\bibfield{author}{\bibinfo{person}{Shahrokh Shahrivar}, \bibinfo{person}{Shaban Elahi}, \bibinfo{person}{Alireza Hassanzadeh}, {and} \bibinfo{person}{Gholamali Montazer}.} \bibinfo{year}{2018}\natexlab{}.
\newblock \showarticletitle{A business model for commercial open source software: A systematic literature review}.
\newblock \bibinfo{journal}{\emph{Information and Software Technology}}  \bibinfo{volume}{103} (\bibinfo{year}{2018}), \bibinfo{pages}{202--214}.
\newblock
\showISSN{0950-5849}
\urldef\tempurl%
\url{https://doi.org/10.1016/j.infsof.2018.06.018}
\showDOI{\tempurl}


\bibitem[Shatalova et~al\mbox{.}(2015)]%
        {shatalova2015methodological}
\bibfield{author}{\bibinfo{person}{Tatyana~N Shatalova}, \bibinfo{person}{Marina~V Chebykina}, \bibinfo{person}{Tatyana~V Zhirnova}, {and} \bibinfo{person}{Elena~Yu Bobkova}.} \bibinfo{year}{2015}\natexlab{}.
\newblock \showarticletitle{Methodological problems in determining the basic features of the sample set controlling the activities of the enterprise}.
\newblock \bibinfo{journal}{\emph{Mediterranean Journal of Social Sciences}} \bibinfo{volume}{6}, \bibinfo{number}{3 S4} (\bibinfo{year}{2015}), \bibinfo{pages}{261}.
\newblock


\bibitem[Spijkerman and Jansen(2018)]%
        {SpijkermanJansen-184}
\bibfield{author}{\bibinfo{person}{Zeena Spijkerman} {and} \bibinfo{person}{Slinger Jansen}.} \bibinfo{year}{2018}\natexlab{}.
\newblock \showarticletitle{The open source software business model blueprint: A comparative analysis of 10 open source companies.}. In \bibinfo{booktitle}{\emph{SiBW}} \emph{(\bibinfo{series}{SiBW})}. \bibinfo{pages}{128--143}.
\newblock


\bibitem[Stallman(2003)]%
        {stallman2003free}
\bibfield{author}{\bibinfo{person}{Richard Stallman}.} \bibinfo{year}{2003}\natexlab{}.
\newblock \showarticletitle{Free software foundation (fsf)}.
\newblock In \bibinfo{booktitle}{\emph{Encyclopedia of Computer Science}}. \bibinfo{pages}{732--733}.
\newblock


\bibitem[Stam(2009)]%
        {Stam-80}
\bibfield{author}{\bibinfo{person}{Wouter Stam}.} \bibinfo{year}{2009}\natexlab{}.
\newblock \showarticletitle{When does community participation enhance the performance of open source software companies?}
\newblock \bibinfo{journal}{\emph{Research Policy}} \bibinfo{volume}{38}, \bibinfo{number}{8} (\bibinfo{year}{2009}), \bibinfo{pages}{1288--1299}.
\newblock


\bibitem[Steinmacher et~al\mbox{.}(2015)]%
        {steinmacher2015systematic}
\bibfield{author}{\bibinfo{person}{Igor Steinmacher}, \bibinfo{person}{Marco Aurelio~Graciotto Silva}, \bibinfo{person}{Marco~Aurelio Gerosa}, {and} \bibinfo{person}{David~F Redmiles}.} \bibinfo{year}{2015}\natexlab{}.
\newblock \showarticletitle{A systematic literature review on the barriers faced by newcomers to open source software projects}.
\newblock \bibinfo{journal}{\emph{Information and Software Technology}}  \bibinfo{volume}{59} (\bibinfo{year}{2015}), \bibinfo{pages}{67--85}.
\newblock


\bibitem[Stewart et~al\mbox{.}(2006)]%
        {StewartAmmeter-131}
\bibfield{author}{\bibinfo{person}{Katherine~J. Stewart}, \bibinfo{person}{Anthony~P. Ammeter}, {and} \bibinfo{person}{Likoebe~M. Maruping}.} \bibinfo{year}{2006}\natexlab{}.
\newblock \showarticletitle{Impacts of License Choice and Organizational Sponsorship on User Interest and Development Activity in Open Source Software Projects}.
\newblock \bibinfo{journal}{\emph{Information Systems Research}} \bibinfo{volume}{17}, \bibinfo{number}{2} (\bibinfo{year}{2006}), \bibinfo{pages}{126--144}.
\newblock
\urldef\tempurl%
\url{https://doi.org/10.1287/isre.1060.0082}
\showDOI{\tempurl}
\newblock
\shownote{identifier: 10.1287/isre.1060.0082}.


\bibitem[Stewart and Gosain(2006)]%
        {StewartGosain-155}
\bibfield{author}{\bibinfo{person}{Katherine~J. Stewart} {and} \bibinfo{person}{Sanjay Gosain}.} \bibinfo{year}{2006}\natexlab{}.
\newblock \showarticletitle{The Impact of Ideology on Effectiveness in Open Source Software Development Teams}.
\newblock \bibinfo{journal}{\emph{MIS quarterly}} \bibinfo{volume}{30}, \bibinfo{number}{2} (\bibinfo{year}{2006}), \bibinfo{pages}{291--314}.
\newblock
\urldef\tempurl%
\url{https://doi.org/10.2307/25148732}
\showDOI{\tempurl}


\bibitem[Tansho and Noda(2015)]%
        {TanshoNoda-97}
\bibfield{author}{\bibinfo{person}{Terutaka Tansho} {and} \bibinfo{person}{Tetsuo Noda}.} \bibinfo{year}{2015}\natexlab{}.
\newblock \showarticletitle{Utilization and development contribution of open source software in Japanese IT companies: an exploratory study of the effect on business growth (2nd report based on 2014 survey)}. In \bibinfo{booktitle}{\emph{Proceedings of the 11th International Symposium on Open Collaboration}} \emph{(\bibinfo{series}{Proceedings of the 11th International Symposium on Open Collaboration})}. \bibinfo{publisher}{Association for Computing Machinery}, \bibinfo{address}{San Francisco, California}, \bibinfo{pages}{Article 3}.
\newblock


\bibitem[Terje(2013)]%
        {snarby2013collaboration}
\bibfield{author}{\bibinfo{person}{Snarby Terje}.} \bibinfo{year}{2013}\natexlab{}.
\newblock \emph{\bibinfo{title}{Collaboration Patterns among Commercial Firms in Community-Based OSS Projects}}.
\newblock \bibinfo{thesistype}{Master's\ thesis}. \bibinfo{school}{Institutt for datateknikk og informasjonsvitenskap}.
\newblock


\bibitem[Van~der Linden et~al\mbox{.}(2009)]%
        {van2009commodification}
\bibfield{author}{\bibinfo{person}{Frank Van~der Linden}, \bibinfo{person}{Bj{\"o}rn Lundell}, {and} \bibinfo{person}{Pentti Marttiin}.} \bibinfo{year}{2009}\natexlab{}.
\newblock \showarticletitle{Commodification of industrial software: A case for open source}.
\newblock \bibinfo{journal}{\emph{IEEE software}} \bibinfo{volume}{26}, \bibinfo{number}{4} (\bibinfo{year}{2009}), \bibinfo{pages}{77--83}.
\newblock


\bibitem[Von~Krogh and Von~Hippel(2006)]%
        {von2006promise}
\bibfield{author}{\bibinfo{person}{Georg Von~Krogh} {and} \bibinfo{person}{Eric Von~Hippel}.} \bibinfo{year}{2006}\natexlab{}.
\newblock \showarticletitle{The promise of research on open source software}.
\newblock \bibinfo{journal}{\emph{Management science}} \bibinfo{volume}{52}, \bibinfo{number}{7} (\bibinfo{year}{2006}), \bibinfo{pages}{975--983}.
\newblock


\bibitem[Wagstrom et~al\mbox{.}(2010)]%
        {wagstrom2010impact}
\bibfield{author}{\bibinfo{person}{Patrick Wagstrom}, \bibinfo{person}{James~D Herbsleb}, \bibinfo{person}{Robert~E Kraut}, {and} \bibinfo{person}{Audris Mockus}.} \bibinfo{year}{2010}\natexlab{}.
\newblock \showarticletitle{The impact of commercial organizations on volunteer participation in an online community}. In \bibinfo{booktitle}{\emph{Academy of Management Annual Meeting}}. \bibinfo{pages}{1--40}.
\newblock


\bibitem[Watson et~al\mbox{.}(2008)]%
        {watson2008business}
\bibfield{author}{\bibinfo{person}{Richard~T Watson}, \bibinfo{person}{Marie-Claude Boudreau}, \bibinfo{person}{Paul~T York}, \bibinfo{person}{Martina~E Greiner}, {and} \bibinfo{person}{Donald Wynn~Jr}.} \bibinfo{year}{2008}\natexlab{}.
\newblock \showarticletitle{The business of open source}.
\newblock \bibinfo{journal}{\emph{Commun. ACM}} \bibinfo{volume}{51}, \bibinfo{number}{4} (\bibinfo{year}{2008}), \bibinfo{pages}{41--46}.
\newblock


\bibitem[Weiss(2018)]%
        {Weiss-106}
\bibfield{author}{\bibinfo{person}{Michael Weiss}.} \bibinfo{year}{2018}\natexlab{}.
\newblock \showarticletitle{Business of Open Source: A Case Study of Integrating Existing Patterns Through Narratives}. In \bibinfo{booktitle}{\emph{Proceedings of the 23rd European Conference on Pattern Languages of Programs}} \emph{(\bibinfo{series}{Proceedings of the 23rd European Conference on Pattern Languages of Programs})}. \bibinfo{publisher}{ACM}, \bibinfo{address}{Irsee, Germany}, \bibinfo{pages}{1--4}.
\newblock


\bibitem[West(2003)]%
        {West-33}
\bibfield{author}{\bibinfo{person}{Joel West}.} \bibinfo{year}{2003}\natexlab{}.
\newblock \showarticletitle{How open is open enough?}
\newblock \bibinfo{journal}{\emph{Research Policy}} \bibinfo{volume}{32}, \bibinfo{number}{7} (\bibinfo{year}{2003}), \bibinfo{pages}{1259--1285}.
\newblock


\bibitem[West and Gallagher(2006a)]%
        {WestGallagher-29}
\bibfield{author}{\bibinfo{person}{Joel West} {and} \bibinfo{person}{Scott Gallagher}.} \bibinfo{year}{2006}\natexlab{a}.
\newblock \showarticletitle{Challenges of open innovation: the paradox of firm investment in open-source software}.
\newblock \bibinfo{journal}{\emph{R and D Management}} \bibinfo{volume}{36}, \bibinfo{number}{3} (\bibinfo{year}{2006}), \bibinfo{pages}{319--331}.
\newblock
\urldef\tempurl%
\url{https://doi.org/10.1111/j.1467-9310.2006.00436.x}
\showDOI{\tempurl}
\newblock
\shownote{identifier: 10.1111/j.1467-9310.2006.00436.x}.


\bibitem[West and Gallagher(2006b)]%
        {west2006challenges}
\bibfield{author}{\bibinfo{person}{Joel West} {and} \bibinfo{person}{Scott Gallagher}.} \bibinfo{year}{2006}\natexlab{b}.
\newblock \showarticletitle{Challenges of open innovation: the paradox of firm investment in open-source software}.
\newblock \bibinfo{journal}{\emph{R\&d Management}} \bibinfo{volume}{36}, \bibinfo{number}{3} (\bibinfo{year}{2006}), \bibinfo{pages}{319--331}.
\newblock


\bibitem[West and O’Mahony(2008)]%
        {west2008designing}
\bibfield{author}{\bibinfo{person}{J West} {and} \bibinfo{person}{S O’Mahony}.} \bibinfo{year}{2008}\natexlab{}.
\newblock \showarticletitle{Designing a participation architecture to support firm-community collaboration}.
\newblock \bibinfo{journal}{\emph{Industry and Innovation}} \bibinfo{volume}{15}, \bibinfo{number}{2} (\bibinfo{year}{2008}), \bibinfo{pages}{145--168}.
\newblock


\bibitem[Williams(2010)]%
        {williams2010free}
\bibfield{author}{\bibinfo{person}{Sam Williams}.} \bibinfo{year}{2010}\natexlab{}.
\newblock \bibinfo{booktitle}{\emph{Free as in freedom (2.0): Richard Stallman and the free software revolution}}.
\newblock \bibinfo{publisher}{Free Software Foundation Boston}.
\newblock


\bibitem[Wohlin(2014)]%
        {wohlin2014guidelines}
\bibfield{author}{\bibinfo{person}{Claes Wohlin}.} \bibinfo{year}{2014}\natexlab{}.
\newblock \showarticletitle{Guidelines for snowballing in systematic literature studies and a replication in software engineering}. In \bibinfo{booktitle}{\emph{Proceedings of the 18th international conference on evaluation and assessment in software engineering}}. \bibinfo{pages}{1--10}.
\newblock


\bibitem[Y. et~al\mbox{.}(2018)]%
        {Y.X.-195}
\bibfield{author}{\bibinfo{person}{Zhang Y.}, \bibinfo{person}{Tan X.}, \bibinfo{person}{Zhou M.}, {and} \bibinfo{person}{Jin Z.}} \bibinfo{year}{2018}\natexlab{}.
\newblock \showarticletitle{Poster: Companies' Domination in FLOSS Development - An Empirical Study of OpenStack}. In \bibinfo{booktitle}{\emph{2018 IEEE/ACM 40th International Conference on Software Engineering: Companion (ICSE-Companion)}} \emph{(\bibinfo{series}{2018 IEEE/ACM 40th International Conference on Software Engineering: Companion (ICSE-Companion)})}. \bibinfo{publisher}{Association for Computing Machinery}, \bibinfo{address}{Gothenburg, Sweden}, \bibinfo{pages}{440--441}.
\newblock


\bibitem[Yang et~al\mbox{.}(2022)]%
        {yang2022survey}
\bibfield{author}{\bibinfo{person}{Yanming Yang}, \bibinfo{person}{Xin Xia}, \bibinfo{person}{David Lo}, {and} \bibinfo{person}{John Grundy}.} \bibinfo{year}{2022}\natexlab{}.
\newblock \showarticletitle{A survey on deep learning for software engineering}.
\newblock \bibinfo{journal}{\emph{ACM Computing Surveys (CSUR)}} \bibinfo{volume}{54}, \bibinfo{number}{10s} (\bibinfo{year}{2022}), \bibinfo{pages}{1--73}.
\newblock


\bibitem[Yu(2020)]%
        {yu2020role}
\bibfield{author}{\bibinfo{person}{Yiqing Yu}.} \bibinfo{year}{2020}\natexlab{}.
\newblock \showarticletitle{Role of reciprocity in firms' open source strategies}.
\newblock \bibinfo{journal}{\emph{Baltic Journal of Management}} \bibinfo{volume}{15}, \bibinfo{number}{5} (\bibinfo{year}{2020}), \bibinfo{pages}{797--815}.
\newblock


\bibitem[Zhang et~al\mbox{.}(2022a)]%
        {Y.H.-30}
\bibfield{author}{\bibinfo{person}{Yuxia Zhang}, \bibinfo{person}{Hao He}, {and} \bibinfo{person}{Minghui Zhou}.} \bibinfo{year}{2022}\natexlab{a}.
\newblock \showarticletitle{Commercial participation in OpenStack: Two sides of a coin}.
\newblock \bibinfo{journal}{\emph{Computer}} \bibinfo{volume}{55}, \bibinfo{number}{2} (\bibinfo{year}{2022}), \bibinfo{pages}{78--84}.
\newblock


\bibitem[Zhang et~al\mbox{.}(2022b)]%
        {ZhangLiu-191}
\bibfield{author}{\bibinfo{person}{Yuxia Zhang}, \bibinfo{person}{Hui Liu}, \bibinfo{person}{Xin Tan}, \bibinfo{person}{Minghui Zhou}, \bibinfo{person}{Zhi Jin}, {and} \bibinfo{person}{Jiaxin Zhu}.} \bibinfo{year}{2022}\natexlab{b}.
\newblock \showarticletitle{Turnover of companies in OpenStack: Prevalence and rationale}.
\newblock \bibinfo{journal}{\emph{ACM Transactions on Software Engineering and Methodology (TOSEM)}} \bibinfo{volume}{31}, \bibinfo{number}{4} (\bibinfo{year}{2022}), \bibinfo{pages}{1--24}.
\newblock


\bibitem[Zhang et~al\mbox{.}(2022c)]%
        {ZhangStol-196}
\bibfield{author}{\bibinfo{person}{Yuxia Zhang}, \bibinfo{person}{Klaas-Jan Stol}, \bibinfo{person}{Hui Liu}, {and} \bibinfo{person}{Minghui Zhou}.} \bibinfo{year}{2022}\natexlab{c}.
\newblock \showarticletitle{Corporate dominance in open source ecosystems: a case study of OpenStack}. In \bibinfo{booktitle}{\emph{Proceedings of the 30th ACM Joint European Software Engineering Conference and Symposium on the Foundations of Software Engineering}} \emph{(\bibinfo{series}{Proceedings of the 30th ACM Joint European Software Engineering Conference and Symposium on the Foundations of Software Engineering})}. \bibinfo{publisher}{Association for Computing Machinery}, \bibinfo{address}{Singapore, Singapore}, \bibinfo{pages}{1048–1060}.
\newblock


\bibitem[Zhang et~al\mbox{.}(2018)]%
        {zhang2018companies}
\bibfield{author}{\bibinfo{person}{Yuxia Zhang}, \bibinfo{person}{Xin Tan}, \bibinfo{person}{Minghui Zhou}, {and} \bibinfo{person}{Zhi Jin}.} \bibinfo{year}{2018}\natexlab{}.
\newblock \showarticletitle{Companies' domination in FLOSS development: an empirical study of OpenStack}. In \bibinfo{booktitle}{\emph{Proceedings of the 40th International Conference on Software Engineering: Companion Proceeedings}}. ACM, \bibinfo{pages}{440--441}.
\newblock


\bibitem[Zhang et~al\mbox{.}(2019)]%
        {zhang2019companies}
\bibfield{author}{\bibinfo{person}{Yuxia Zhang}, \bibinfo{person}{Minghui Zhou}, \bibinfo{person}{Audris Mockus}, {and} \bibinfo{person}{Zhi Jin}.} \bibinfo{year}{2019}\natexlab{}.
\newblock \showarticletitle{Companies’ participation in oss development--an empirical study of openstack}.
\newblock \bibinfo{journal}{\emph{IEEE Transactions on Software Engineering}} \bibinfo{volume}{47}, \bibinfo{number}{10} (\bibinfo{year}{2019}), \bibinfo{pages}{2242--2259}.
\newblock


\bibitem[Zhang et~al\mbox{.}(2020)]%
        {Y.M.-190}
\bibfield{author}{\bibinfo{person}{Yuxia Zhang}, \bibinfo{person}{Minghui Zhou}, \bibinfo{person}{Klaas-Jan Stol}, \bibinfo{person}{Jianyu Wu}, {and} \bibinfo{person}{Zhi Jin}.} \bibinfo{year}{2020}\natexlab{}.
\newblock \showarticletitle{How do companies collaborate in open source ecosystems? an empirical study of openstack}. In \bibinfo{booktitle}{\emph{Proceedings of the ACM/IEEE 42nd International Conference on Software Engineering}}. \bibinfo{pages}{1196--1208}.
\newblock


\bibitem[Zhang et~al\mbox{.}(2017)]%
        {yuxia2017openstack}
\bibfield{author}{\bibinfo{person}{Yuxia Zhang}, \bibinfo{person}{Minghui Zhou}, \bibinfo{person}{Wei Zhang}, \bibinfo{person}{Hanyan Zhao}, {and} \bibinfo{person}{Zhi Jin}.} \bibinfo{year}{2017}\natexlab{}.
\newblock \showarticletitle{How Commercial Organizations Participate in OpenStack Open Source Projects}.
\newblock \bibinfo{journal}{\emph{Journal of Software}} \bibinfo{volume}{28}, \bibinfo{number}{6} (\bibinfo{year}{2017}), \bibinfo{pages}{1343--1356}.
\newblock


\bibitem[Zhou et~al\mbox{.}(2016a)]%
        {ZhouMockus-41}
\bibfield{author}{\bibinfo{person}{Minghui Zhou}, \bibinfo{person}{Audris Mockus}, \bibinfo{person}{Xiujuan Ma}, \bibinfo{person}{Lu Zhang}, {and} \bibinfo{person}{Hong Mei}.} \bibinfo{year}{2016}\natexlab{a}.
\newblock \showarticletitle{Inflow and Retention in OSS Communities with Commercial Involvement}.
\newblock \bibinfo{journal}{\emph{ACM Transactions on Software Engineering and Methodology}} \bibinfo{volume}{25}, \bibinfo{number}{2} (\bibinfo{year}{2016}), \bibinfo{pages}{1--29}.
\newblock
\urldef\tempurl%
\url{https://doi.org/10.1145/2876443}
\showDOI{\tempurl}
\newblock
\shownote{identifier: 10.1145/2876443}.


\bibitem[Zhou et~al\mbox{.}(2016b)]%
        {zhou2016inflow}
\bibfield{author}{\bibinfo{person}{Minghui Zhou}, \bibinfo{person}{Audris Mockus}, \bibinfo{person}{Xiujuan Ma}, \bibinfo{person}{Lu Zhang}, {and} \bibinfo{person}{Hong Mei}.} \bibinfo{year}{2016}\natexlab{b}.
\newblock \showarticletitle{Inflow and retention in oss communities with commercial involvement: A case study of three hybrid projects}.
\newblock \bibinfo{journal}{\emph{ACM Transactions on Software Engineering and Methodology (TOSEM)}} \bibinfo{volume}{25}, \bibinfo{number}{2} (\bibinfo{year}{2016}), \bibinfo{pages}{13}.
\newblock


\bibitem[Ågerfalk and Fitzgerald(2008)]%
        {agerfalkFitzgerald-180}
\bibfield{author}{\bibinfo{person}{Pär~J. Ågerfalk} {and} \bibinfo{person}{Brian Fitzgerald}.} \bibinfo{year}{2008}\natexlab{}.
\newblock \showarticletitle{Outsourcing to an Unknown Workforce: Exploring Opensurcing as a Global Sourcing Strategy}.
\newblock \bibinfo{journal}{\emph{MIS quarterly}} \bibinfo{volume}{32}, \bibinfo{number}{2} (\bibinfo{year}{2008}), \bibinfo{pages}{385--409}.
\newblock
\urldef\tempurl%
\url{https://doi.org/10.2307/25148845}
\showDOI{\tempurl}


\end{thebibliography}

\end{document}